\documentclass[acmsmall]{acmart}
\usepackage{xspace}
\usepackage{multirow}
\usepackage{listings}
\usepackage{xcolor}
\usepackage{minted}
\usepackage[utf8]{inputenc}
\usepackage[most]{tcolorbox}
\tcbuselibrary{skins, breakable}
\usepackage{xcolor}
\definecolor{lb}{HTML}{E6E6E6} 
\usepackage[table]{xcolor}
\definecolor{promptgray}{RGB}{245,245,245}
\definecolor{promptframe}{RGB}{100,100,100}
\usepackage{subcaption}
\tcbuselibrary{skins}
\tcbuselibrary{minted}
\usepackage{stackengine}
\usepackage[ruled,vlined]{algorithm2e}

\newcommand{\ourmethod}{\textsc{AdverTest}\xspace}
\AtBeginDocument{%
  }

\setcopyright{cc}
\setcctype{by-nc-nd}
\acmDOI{10.1145/3832122}
\acmYear{2026}
\acmJournal{PACMSE}
\acmVolume{3}
\acmNumber{ISSTA}
\acmArticle{ISSTA031}
\acmMonth{10}
\acmSubmissionID{issta26main-p285-p}
\received{2026-01-30}
\received[accepted]{2026-06-25}

\usepackage{booktabs}
\usepackage{caption}
\usepackage{textcomp}
\usepackage{enumitem}

\definecolor{RevisionRed}{HTML}{C81E1E}
\newcommand{\revtext}[1]{{#1}}
\definecolor{ShepherdBlue}{HTML}{1A56DB}
\newcommand{\newrev}[1]{#1}   

\colorlet{goodbg}{green!30!white}   
\colorlet{badbg} {red!25!white} 
\definecolor{slicebg}{HTML}{E8F8FF}
\newtcblisting{CodeBox}[2][]{ breakable,
  listing engine   = minted,          
  minted language  = Java,               
  minted options   = {                 
    linenos,                            
    breaklines,                        
    fontsize=\small
  },
  colback         = white,
  colframe        = black!30,
  arc             = 2pt,
  outer arc       = 2pt,
  breakable,
  left=3pt,right=3pt,top=2pt,bottom=2pt,
  #1                                   
}
\newtcolorbox{answer2RQ}[2][]{%
  enhanced,
  sharp corners=all,
  colback=white,
  colframe=black!80,
  boxrule=1pt,
  arc=0mm,
  outer arc=0mm,
  left=1mm,
  right=1mm,
  top=1mm,
  bottom=1mm,
  fonttitle=\bfseries,
  coltitle=black,
  colbacktitle=gray!20,
  before skip=1.5ex,
  after skip=1.5ex,
  title={Answer to RQ~#2:},
  #1 
}
\begin{document}

\title{Test vs Mutant: Adversarial LLM Agents for Robust Unit Test Generation}

\newcommand{\sjtucompletednote}{This work was completed while the authors were studying at Shanghai Jiao Tong University.}

\author{Pengyu Chang}
\authornote{\sjtucompletednote}
\orcid{0009-0005-8323-5829}
\affiliation{%
  \institution{Shanghai Jiao Tong University}
  \city{Shanghai}
  \country{China}
}
\email{pengyuch@andrew.cmu.edu}

\author{Yixiong Fang}
\authornotemark[1]
\orcid{0009-0004-2483-8622}
\affiliation{%
  \institution{Shanghai Jiao Tong University}
  \city{Shanghai}
  \country{China}
}
\email{yixiongf@andrew.cmu.edu}

\author{Silin Chen}
\orcid{0009-0001-9914-4172}
\affiliation{%
  \institution{Shanghai Jiao Tong University}
  \city{Shanghai}
  \country{China}
}
\email{csl2457029646@163.com}

\author{Yuling Shi}
\orcid{0009-0009-9738-7072}
\affiliation{%
  \institution{Shanghai Jiao Tong University}
  \city{Shanghai}
  \country{China}
}
\email{yuling.shi@sjtu.edu.cn}

\author{Beijun Shen}
\orcid{0000-0001-8370-3956}
\affiliation{%
  \institution{Shanghai Jiao Tong University}
  \city{Shanghai}
  \country{China}
}
\email{bjshen@sjtu.edu.cn}

\author{Xiaodong Gu}
\orcid{0000-0002-0529-6408}
\correspondingauthor
\affiliation{%
  \institution{Shanghai Jiao Tong University}
  \city{Shanghai}
  \country{China}
}
\email{xiaodong.gu@sjtu.edu.cn}

\renewcommand{\shortauthors}{Pengyu Chang, Yixiong Fang, Silin Chen, Yuling Shi, Beijun Shen, and Xiaodong Gu}

\begin{abstract}
  Software testing is a critical, yet resource-intensive phase of the software development lifecycle. Search-based approaches typically achieve high coverage but produce tests with low readability, whereas large language model (LLM)-based methods generate more human-readable tests but often suffer from low coverage and compilability. \revtext{While the majority of research efforts have focused on improving test coverage and readability, comparatively less attention has been paid to enhancing the robustness of bug detection.}
  To address this gap, we propose \ourmethod, a novel adversarial framework for LLM-powered test case generation that pairs a test case generation agent ($\mathcal{T}$) with a mutant generation agent ($\mathcal{M}$): $\mathcal{M}$ persistently creates mutants ``hacking'' the blind spots of $\mathcal{T}$'s current test suite, while $\mathcal{T}$ iteratively refines its tests to ``kill'' the challenging mutants, with the interaction guided by both coverage and mutation scores. Experimental results on Defects4J show that our approach improves fault detection rates by 8.56\% over the best existing LLM-based methods and by \revtext{50.20\%} over EvoSuite, while \revtext{remaining competitive on} line and branch coverage.
\end{abstract}



\maketitle

\section{Introduction}

Unit testing is a critical and resource-intensive phase in the software development lifecycle, forming the foundation for building robust software. However, writing high-quality unit tests remains a tedious and time-consuming task for developers~\cite{beller2015and,beller2015much}. The goal of automated test case generation is to alleviate this burden by generating high-quality test cases that can cover diverse program behaviors and detect faults efficiently. \revtext{Throughout this paper, we use \emph{robustness} to mean a test suite's ability to expose diverse fault behaviors, captured by our primary metric, the fault detection rate (FDR).}

There have been various approaches for automated test case generation, such as random testing~\cite{Pacheco2007}, symbolic execution~\cite{PeX,KLEE,maciver2019hypothesis}, property-based testing~\cite{QUICKCHECK,KORAT}, and search-based software testing (SBST) \cite{evosuite,pynguin}. \revtext{While these traditional methods can achieve substantial code coverage, the resulting tests are often less readable and harder to maintain~\cite{deljouyi2024leveraging}.} As a result, they can increase developer effort for debugging and comprehension, as well as generate too few assertions for effective fault detection.

In recent years, the growing capabilities of LLMs to generate and understand human-readable code ~\cite{chen2025swe,lin2024llms,202605.2065,li2025swe,shi2025longcodezip,wang2026swe,hu2026line,zeng2026glimprouter} have provided new opportunities for automated test case generation. For instance, UTGen \cite{deljouyi2024leveraging} integrates LLMs into the SBST process, yielding tests that are both effective and understandable. Similarly, CodaMosa \cite{Lemieux2023} addresses the fitness plateau problem in search-based testing by incorporating LLMs into the test generation process. As a result, CodaMosa outperforms its baseline methods, such as Pynguin~\cite{pynguin} and Codex~\cite{chen2021codex}, in terms of code coverage. Additionally, HITS \cite{hits} demonstrates LLMs' ability to generate high-coverage tests for complex methods by generating tests slice by slice.

\revtext{However, much of this work evaluates generated tests primarily using code coverage metrics; fewer studies explicitly optimize for bug-detection capability, particularly robustness against edge cases or boundary conditions.} It is widely acknowledged that high coverage does not necessarily equate to strong fault detection~\cite{cai2005effect,gopinath2014code,hemmati2015effective}. Recent LLM-based studies typically rely on environmental feedback to iteratively improve their test suites. However, most of these methods use only compiler error messages or coverage metrics for prompt refinement \cite{chatunitest,testforge}. This approach often overlooks logical or semantic faults because coverage metrics only quantify the extent of code execution, not whether the correctness of that execution is rigorously verified. Consequently, a high coverage test suite may still fail to distinguish between correct and incorrect program behaviors.

To address these gaps, we turn to mutation testing (MT), a white-box testing technique that evaluates the ability of a test suite to detect faults. MT injects artificial faults, called \textit{mutants}, into the program by making slight, grammatically correct changes to the code. The test suite is then executed on both the original program and each mutant, with any test that produces a different outcome on a mutant being counted as having ``killed'' that mutant. The mutation score (MS) is the ratio of killed mutants to the total number of generated mutants.

Building on these insights, we propose \textbf{\ourmethod}, a \textbf{Mutation-guided}, \textbf{Adversarial}, \textbf{LLM-driven}, \textbf{Dual-agent} unit test generation framework designed to enhance bug detection capabilities. \revtext{\ourmethod pairs a \textit{Test Case Generation Agent} (\(\mathcal{T}\)), which writes tests to detect bugs, with a \textit{Mutant Generation Agent} (\(\mathcal{M}\)), which produces mutants that evade \(\mathcal{T}\)'s current suite. The two agents co-evolve through a bidirectional feedback loop guided by mutation score and coverage, exposing each other's blind spots over successive rounds; we detail the loop in Section~\ref{sec:meth_adviter}.}

We evaluate \ourmethod on real-world Java projects from Defects4J~\cite{defects4j} and GrowingBugs~\cite{GrowingBugsICSE21,GrowingBugsTSE2022,NaturalnessOfBugsFSE2022}, \revtext{and additionally on 12 Python issues from SWE-Rebench~\cite{badertdinov2026swe} to verify cross-language generalization}. Compared with state-of-the-art baselines (HITS~\cite{hits}, ChatUniTest~\cite{chatunitest}, EvoSuite~\cite{evosuite}\revtext{, and Randoop~\cite{Pacheco2007}}), \ourmethod significantly improves fault detection while \revtext{remaining competitive on} line and branch coverage\revtext{, and the same adversarial loop carries over to Python with 11/12 detection}. Ablation studies further isolate the contribution of the adversarial iteration loop and the mutant-aware feedback mechanism\revtext{; complementary analyses examine the effect of iteration rounds and the choice of backbone LLM}.
\clearpage
In summary, our key contributions are as follows:
\begin{itemize}
  \item We propose \ourmethod, an adversarial dual-agent framework in which two LLM-driven agents generate mutants and test cases with bidirectional feedback on mutation scores and coverage.
  \item \revtext{We evaluate \ourmethod on real-world projects in both Java (Defects4J~\cite{defects4j} and GrowingBugs~\cite{GrowingBugsICSE21,GrowingBugsTSE2022,NaturalnessOfBugsFSE2022}) and Python (SWE-Rebench~\cite{badertdinov2026swe}), demonstrating an 8.56\% gain in fault detection over the strongest LLM baseline and a 50.20\% gain over the search-based baseline, while remaining competitive on line and branch coverage.}
  \item We make \ourmethod publicly available online~\cite{replication} to facilitate replication and future extensions.
\end{itemize}

\section{Related Works}
\subsection{Automated Test Case Generation}
Automated unit test generation has progressed from early heuristic approaches to advanced AI-driven methods.
Traditional techniques such as feedback-directed random testing (e.g., Randoop \cite{Pacheco2007}) and search-based tools (e.g., EvoSuite \cite{evosuite}) achieve high code coverage but often produce tests that are hard to maintain or understand.
To improve quality, researchers reframed test generation as a code synthesis task solvable with machine learning.
For example, Tufano \textit{et al.}\ trained a transformer model on code–test pairs (AthenaTest) to automatically generate JUnit tests for a given method \cite{Tufano2020}.
Subsequent neural approaches introduced refinements: A3Test added assertion knowledge and naming consistency checks to improve correctness \cite{Alagarsamy2024}, and other systems (e.g., ConTest, TeCo, CAT-LM) enhanced semantic understanding and output readability \cite{Villmow2021, Nie2023, Rao2023}.

The emergence of LLMs has further accelerated progress.
Empirical studies showed that modern code-generating LLMs (e.g., Codex or GPT-3.5) can produce unit tests in a human-like style~\cite{shi2025between}, but they struggled to achieve high coverage on complex code and often introduced ``test smells'' (redundant or trivial tests) \cite{Siddiqa2023}.
To better harness LLMs, researchers have integrated these models with program analysis and feedback.
For instance, tools like \textsc{TestPilot} and \textsc{ChatUniTest} pair LLM-based generation with static analysis and verification loops to produce valid, high-coverage tests \cite{Schafer2023, chatunitest}.
Hybrid strategies have also emerged: for example, \textsc{CODAMOSA} invokes an LLM when a search-based test generator hits a coverage plateau, generating tests for hard-to-cover functionality \cite{Lemieux2023}.
Additionally, some frameworks use an iterative loop where an LLM refines its tests based on feedback from prior test runs (e.g., coverage gaps or errors) \cite{testforge,shi2024code}. \textsc{CoverUp} uses coverage rate as a feedback and achieves a higher coverage than \textsc{CODAMOSA} on most modules \cite{coverup}.
These LLM-driven techniques are yielding test suites that are not only more readable but also more effective at finding bugs, with substantial gains in coverage and fault detection reported in both research and industry ~\cite{Alshahwan2024}.

Unlike prior LLM‑based generators that use one‑way feedback (e.g., compiler errors or coverage gaps) to refine tests, \ourmethod introduces a second LLM that creates context‑aware mutants and engages the test generator in an adversarial loop. This bidirectional loop, guided by mutation‑score and coverage feedback, pushes each agent to close the other’s blind spots and yields more robust fault detection.

\revtext{Co-evolutionary and adversarial dynamics between tests and program variants pre-date the LLM era and inspire our design. Arcuri and Yao~\cite{ArcuriYao2014} proposed a co-evolutionary search loop in which a program and its test suite alternately evolve to overcome each other. Closer to our setting, \emph{Code Defenders}~\cite{CodeDefenders2019} gamifies mutation testing for education by assigning human students to opposing mutant-writing and test-writing teams, mirroring the role split between Agents $\mathcal{M}$ and $\mathcal{T}$. \ourmethod operationalizes this adversarial structure with LLM-driven agents, replacing fixed operator sets and human players with context-aware, feedback-driven generation.}

\begin{figure*}[!tbp]
  \centering
  \includegraphics[width=\linewidth ]{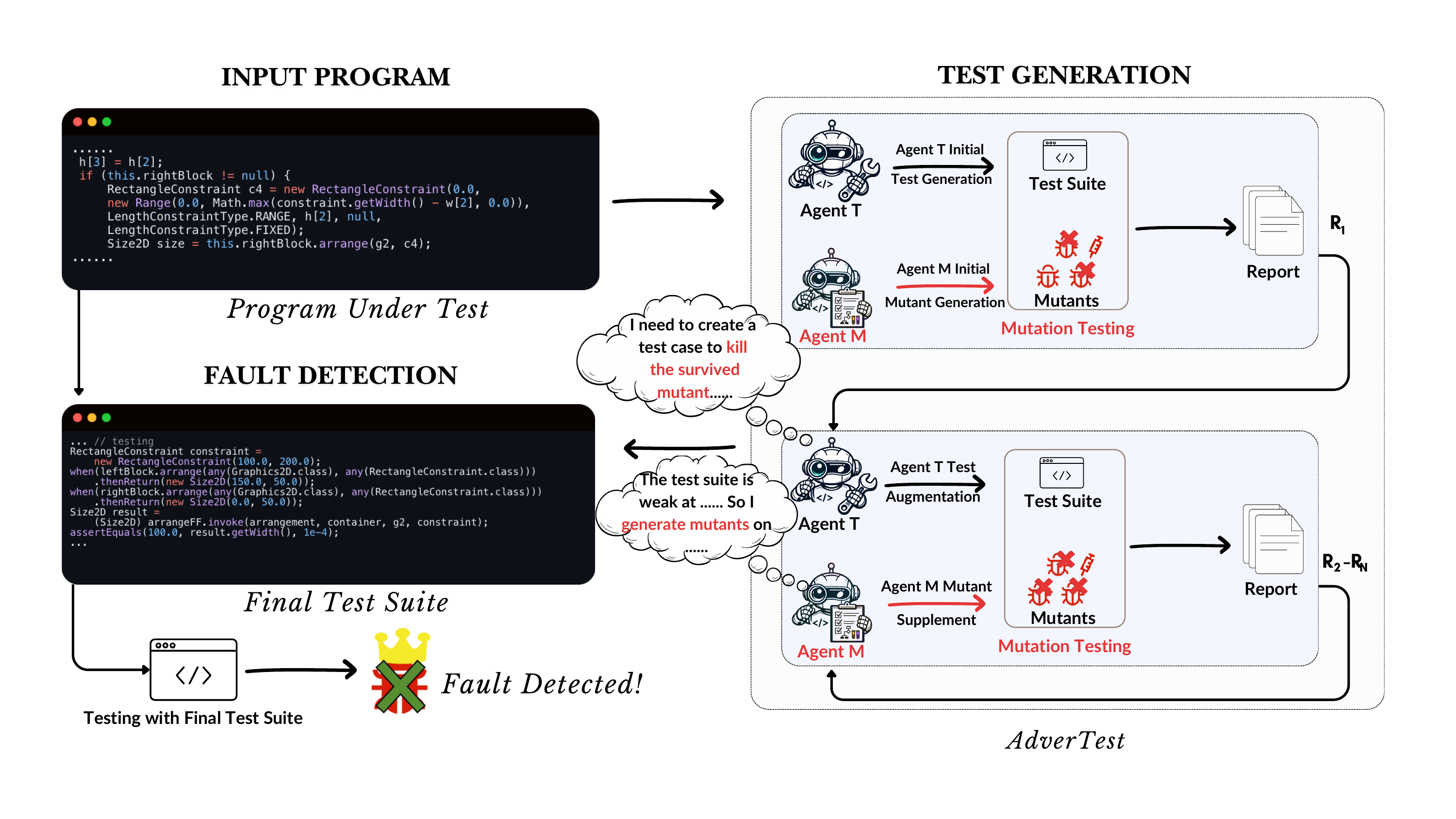}
  \caption{Overview of \ourmethod. Agents T and M alternately generate tests and create mutants, guided by coverage and mutation‐score feedback.}
  \Description{Figure for the framework}
  \label{framework_fig}
\end{figure*}
\subsection{Mutation Testing}
Mutation testing evaluates a test suite’s rigor by seeding artificial faults (mutants) into the program and checking if the tests detect them \cite{DeMillo1978}.
In the classical approach, developers apply simple code modifications (mutation operators) to produce mutants, then run the suite on each mutant; the fraction of mutants causing test failures (the mutation score) indicates the suite’s fault-detection effectiveness \cite{Coles2016, Just2011}.

MT has long been used in traditional automated test generation workflows; for example, EvoSuite~\cite{evosuite} applies mutation testing to synthesize assertions. In LLM-based methods, the effects of MT have not yet been thoroughly explored. MuTAP~\cite{mutap} was the first to explore the generation of LLM-based test cases with mutation testing. MuTAP integrates surviving mutants into LLM prompts to improve fault detection, but was evaluated solely on HumanEval~\cite{chen2021codex} and Refactory~\cite{yang2019refactory}, a dataset of student submitted buggy programs, not on industrial scale projects. Despite this narrow scope, MuTAP achieved notable gains in both mutation score and bug detection rate. Recently, Harman et al. utilized LLM-generated mutants in test generation, but it is not adversarial and evolutionary~\cite{mutationatmeta}. Barboni et al. employ an ML model to evaluate the usefulness of each surviving mutant, then prompt them to LLM to generate test cases for smart contracts~\cite{barboni2025smartmutant}.

Building on this idea, large pre-trained models have been employed to generate mutants without manual rule design.
For instance, $\mu$BERT repurposes a Transformer-based code model (CodeBERT) to suggest likely mutations by predicting masked tokens in code. \textsc{LLMorpheus} prompts an LLM to inject diverse bugs into code \cite{Degiovanni2022, Tip2025}.

Studies have shown that LLM-generated mutants are more diverse and effective at revealing bugs than those from conventional tools: Wang \textit{et al.}\ report that GPT-4 mutants improved real fault detection by nearly 30\% over the best rule-based approach in a benchmark evaluation \cite{WangMutation2024}.
Similarly, LLM-created mutants often mimic real vulnerabilities, breaking the same test cases as the actual faults \cite{Garg2024}.

\ourmethod further utilizes LLM generated mutants in the test case generating process. By replacing fixed mutation operators with an LLM and alternating test and mutant reinforcement, \ourmethod is one of the first frameworks to couple adversarial LLM test generation with LLM mutant generation, outperforming both search‑based tools and earlier LLM methods on real-world defects.

\section{Methodology}

In this section, we introduce \textbf{\ourmethod}, an adversarial mutation-guided unit test generation framework in detail. 

\subsection{Framework Overview}

Figure~\ref{framework_fig} illustrates the overall framework of \ourmethod, which consists of an adversarial loop between a \textit{Test Case Generation Agent} ($\mathcal{T}$) and a \textit{Mutant Generation Agent} ($\mathcal{M}$). These agents iteratively refine the test suite and generate increasingly challenging mutants. The specific mechanisms of these components and the iterative process are detailed in Sections \ref{sec:meth_initest}-\ref{sec:meth_adviter}.

\subsection{Initial Test Suite Generation}
\label{sec:meth_initest}

Given a target program \(P\), the agent $\mathcal{T}$ is instructed to generate an initial test suite \(T\). The generation prompt consists of three components: (1) a high-level instruction (e.g., ``generate unit tests for the following program''), (2) the complete source code of the program under test, and (3) relevant contextual information, such as surrounding method signatures and class constructors. The full prompt template is designed as follows:

\definecolor{PromptBack}{HTML}{F8FAFC}      
\definecolor{PromptFrame}{HTML}{111827}     
\definecolor{PromptTitleBack}{HTML}{050505} 
\definecolor{PromptSep}{HTML}{CBD5E1}       

\definecolor{PromptAccent}{HTML}{1D4ED8}    
\definecolor{PromptVar}{HTML}{B42318}       

\newcommand{\SectionTag}[1]{\textbf{\textcolor{PromptAccent}{[#1]}}}
\newcommand{\Placeholder}[1]{\texttt{\textcolor{PromptVar}{\{#1\}}}}
\newcommand{\PlaceholderInTT}[1]{\textcolor{PromptVar}{\{#1\}}}

\begin{tcolorbox}[
    breakable,
    enhanced,
    title=\textbf{Prompt Template for Initial Test Suite Generation},
    colback=PromptBack,
    colframe=PromptFrame,
    coltitle=white,
    colbacktitle=PromptTitleBack,
    fonttitle=\bfseries,
    fontupper=\footnotesize,
    fontlower=\footnotesize,
    boxrule=0.9pt,
    left=2mm, right=2mm, top=2mm, bottom=2mm,
    arc=0mm,
    titlerule=0pt,
    boxed title style={
        sharp corners,
        colframe=PromptTitleBack,
        colback=PromptTitleBack,
        left=2mm, right=2mm, top=1mm, bottom=1mm
    },
    segmentation style={draw=PromptSep, line width=0.4pt}
]

\SectionTag{Instruction}\\
You are an expert Java developer and software tester. Your task is to generate full JUnit test methods for a given Java method inside a Java class. Follow these steps to ensure comprehensive and effective test coverage:

\begin{enumerate}[leftmargin=*, nosep]
    \item \textbf{Analyze the Java Method}:\\
    ...
    \item \textbf{Design Test Cases}:\\
    ...
    \item \textbf{Implement the Test Method}:\\
    ...
\end{enumerate}

\tcbline 

\SectionTag{Example}\\
(An Example of a Java Method and corresponding Test Case)

\tcbline 

\SectionTag{Task Inputs}\\
Given the following Java method, generate a complete JUnit test method that thoroughly tests the method. Utilize your reasoning ability to ensure that all possible scenarios and edge cases are considered.

\vspace{0.5em}
\textbf{Input Java Method (\Placeholder{method\_name}):}

\texttt{method\_body: \PlaceholderInTT{method\_code}}

\textbf{Class Context:}
\begin{itemize}[leftmargin=*, nosep]
    \item \textbf{Other Class Variables:} \Placeholder{class\_variables}
    \item \textbf{Other Methods in the Class (no method body shown):} \Placeholder{method\_info}
    \item \textbf{Constructors of the class object:} \Placeholder{Constructors}
\end{itemize}

\tcbline 

\SectionTag{Guidelines}\\
(Specific Guidelines for generation, including Java and JUnit version and output format)

\end{tcolorbox}

As an example, we show the prompt template for Agent $\mathcal{T}$'s initial test case generation. In the Instruction part, we assign the agent's specific roles, Java developer and software testing engineer, and give an initial description of their tasks. Then, we apply a chain-of-thought (CoT)\cite{wei2022chain} inspired approach to outline the procedures and guidelines that they should follow. In the Example section, we provide a few examples for a model to learn how to generate test cases. In the Task Inputs section, we provide a detailed task description along with all the necessary contextual information. Finally, the Guidelines section offers targeted guidance, including the exact versions of any required software packages and other relevant details. Our methodology follows an iterative prompt engineering process, where each prompt is tested and refined based on observed results. The targeted guidance in the Guidelines section also incorporates common failure modes previously exhibited during the process, which we manually identified and embedded in the prompt, to significantly reduce the likelihood of repeated errors.

Following previous works of ChatUniTest \cite{chatunitest} and HITS~\cite{hits}, each initial test case undergoes a repair process to ensure that only syntactically valid and runnable tests are included in the generated test suite \(T\). The raw test cases are compiled and executed against the original program \(P\). If the test fails to compile or execute, we apply 6 deterministic rules to fix them (e.g., fixing missing semicolons, balancing braces). 

Having applied all rules, the initial test is recompiled and re-executed. The process stops at the first successful revision. 

If all rule-based attempts fail, we invoke an LLM-guided repair process for up to \(K\) rounds. In each round, an LLM is instructed with a bug fix prompt, consisting of the last candidate test and its corresponding compilation or runtime error messages. The LLM returns a revised version, which is again compiled and executed. The first syntactically correct and passing revision is accepted. 

Following previous works~\cite{hits,chatunitest}, we set \(K = 10\). If a test cannot be repaired successfully within the allocated attempts, it is discarded.

It is worth noting that while all tests in \(T\) are compilable, they may still include runtime failures (e.g., assertion errors) by design, as these are often indicative of bugs in the program under test. Such failure-exposing tests are retained, as they are valuable for driving mutation testing and bug detection in subsequent phases.

\subsection{Initial Mutant Generation}

While Agent \(\mathcal{T}\) generates an initial test suite, Agent \(\mathcal{M}\) works in parallel to generate an initial set of mutants \(M\) for the program under test. 
The goal is to generate diverse and behaviorally meaningful mutants that are both syntactically valid and semantically different from the original program. 

The mutant generation process is conducted in a prompt-driven manner. Specifically, we instruct the LLM with a meticulously designed prompt comprising three components: (1) A natural language instruction of the mutation task, (2) The complete code context of the program \(P\), and (3) A set of mutation examples formatted as JSON objects, each illustrating a valid single-line mutation. 
The few-shot examples are drawn from a previous work by Wang~\textit{et al.}~\cite{WangMutation2024}, which curated mutation examples from the QuixBugs~\cite{Quixbugs} benchmark, which was different from our evaluation dataset while being representative of bugs.

To further promote mutation diversity and correctness, we adopt prior research on prompt-based mutation~\cite{Tip2025,WangMutation2024}. Specifically, we enforce a \textbf{single-line modification} constraint. This design choice is grounded in the two fundamental hypotheses of mutation testing: the \textit{Competent Programmer Hypothesis} and the \textit{Coupling Effect}. The latter asserts that ``test data that distinguishes all programs differing from a correct one by only simple errors is so sensitive that it also distinguishes more complex errors''~\cite{DeMillo1978,OFFUT92COUPLING}. By restricting mutants to single-line changes, we focus on these relatively simple errors, ensuring the generated test suite is sensitive enough to detect complex faults while minimizing the generation of uncompilable code common in unconstrained LLM generation~\cite{WangMutation2024}. 

Accordingly, we provide the following constraints in the instruction:
\begin{itemize}
    \item Only one mutation is allowed per mutant.
    \item Each mutation must modify exactly one line of code.
    \item Redundant or meaningless mutations (e.g., altering comments or whitespace) are disallowed.
    \item Output format is strictly specified to enable parsing and integration into the mutation testing infrastructure.
    \item Previously generated mutants must not be repeated.
\end{itemize}


\definecolor{PromptBack}{HTML}{F8FAFC}
\definecolor{PromptFrame}{HTML}{111827}
\definecolor{PromptTitleBack}{HTML}{050505} 
\definecolor{PromptSep}{HTML}{CBD5E1}
\definecolor{ExampleText}{HTML}{15803D} 
\definecolor{PromptAccent}{HTML}{1D4ED8}    
\definecolor{PromptVar}{HTML}{B42318}       

\begin{tcolorbox}[
    breakable,
    enhanced,
    title=\textbf{Prompt Template for Initial Mutant Generation},
    colback=PromptBack,
    colframe=PromptFrame,
    coltitle=white,
    colbacktitle=PromptTitleBack,
    fonttitle=\bfseries,
    fontupper=\footnotesize,
    fontlower=\footnotesize,
    boxrule=0.9pt,
    left=2mm, right=2mm, top=2mm, bottom=2mm,
    arc=0mm,
    titlerule=0pt,
    boxed title style={
        sharp corners,
        colframe=PromptTitleBack,
        colback=PromptTitleBack,
        left=2mm, right=2mm, top=1mm, bottom=1mm
    },
    segmentation style={draw=PromptSep, line width=0.4pt}
]

\SectionTag{Instruction}\\
Below is a code snippet from a Java project. Your task is to generate \Placeholder{MUT\_NUM} mutants for this code.
\textit{(Note: A mutant refers to a syntactically valid variant with a subtle alteration used for software testing.)}

\vspace{0.5em}
\SectionTag{Input Code}

\texttt{\PlaceholderInTT{code}}


\tcbline 

\SectionTag{Few-Shot Examples}\\
Refer to the following format for the expected output logic:
\begingroup
\color{ExampleText}
\begin{verbatim}
{
  "id": "1",
  "precode": "return depth==0;",
  "aftercode": "return true;",
  "line_number": 42
}
\end{verbatim}
\endgroup

\tcbline 

\SectionTag{Requirements}
\begin{enumerate}[leftmargin=*, nosep]
    \item Provide the generated mutants directly in the specified output format.
    \item \textbf{Single Line Modification:} Ensure each mutant affects \textbf{only one line}. Pay strict attention to line breaks; statements may need to be split.
    \item \textbf{Output Format:} Adhere strictly to the JSON format shown above.
    \item \textbf{Uniqueness:} Do not generate duplicate mutants.
    \item \textbf{No meaningless mutants:} Do not generate mutants that result in identical execution behavior to the original.
\end{enumerate}

\end{tcolorbox}



\subsection{Mutation Testing}
\begin{algorithm}[h]
  \caption{Mutation Testing}
  \small
  \LinesNumbered
  \label{alg:mutation-testing}
  \KwIn{Program $P$, test suite $T$, mutants $M$}
  \KwOut{Surviving mutants $M_s$, coverage $C$, mutation score $S$, valid mutants $M_v$}
  \SetAlgoLined
  $M_s \leftarrow \{\}$\;
  \ForEach{$m$ \(\in\) $M$}{
    \If{$m$ compiles}{
        $M_v \leftarrow M_v \cup \{m\}$\;
      $\Delta \leftarrow \textsc{RunTests}(T, m)$\;  
      \If{$\Delta = \emptyset$}{  
        $M_s \leftarrow M_s \cup \{m\}$\;
      }
    }
  }
  $C \leftarrow \textsc{ComputeCoverage}(T, P)$\;
  $S \leftarrow (|M_v| - |M_s|) / |M_v|$\;  
  \Return $(M_s, C, S, M_v)$\;
\end{algorithm}
Once the initial test suite \(T\) and the mutation set \(M\) have been generated, we perform mutation testing on the program under test \(P\) following Algorithm~\ref{alg:mutation-testing}. 

Let \(P\) denote the program under test, \(M\) be the set of mutants created by Agent~\(\mathcal{M}\), and \(T\) the current test suite produced by Agent~\(\mathcal{T}\).
\revtext{We treat $P$ as the reference (``bug-free'') program throughout the generation loop. Agents $\mathcal{T}$ and $\mathcal{M}$ receive only $P$ as input; any buggy variant of $P$, if available, is used solely \emph{outside} the loop to score fault detection.}
\revtext{For each mutant \(m \in M\), the compile step is the sole execution gate: $m$ is added to the valid set \(M_v\) if and only if it compiles within the sandbox (Algorithm~\ref{alg:mutation-testing}, Line~4). Mutants that fail to compile or exceed a wall-clock timeout are discarded as invalid and excluded from all subsequent metrics. Valid mutants are then executed exactly \emph{once} as part of the suite-evaluation step \textsc{RunTests}$(T, m)$ in Line~5; there is no separate ``validate-then-rerun'' phase.}

A mutant is considered \textbf{survived} if it exhibits no behavioral difference from the original program. Concretely, let Fail(\(T\), \(x\)) denote the set of test cases in \(T\) that fail when executed on program \(x\); a mutant \(m\) survives if:
\[
     \Delta = \text{Fail}(T, m)\setminus\text{Fail}(T,P)=\varnothing
\]
\revtext{The set subtraction is retained even though $P$ is bug-free by construction, because LLM-generated tests can be flaky or contain hallucinated assertions that fail on $P$ itself; subtracting $\text{Fail}(T,P)$ ensures that only failures \emph{caused by the mutation} count toward killing the mutant.}
All surviving mutants are added to the surviving set \(M_s\). Conversely, remaining valid mutants that are detected by at least one test case are considered \textbf{killed}.

This mutation testing process produces two key feedback signals: 
(1) the \emph{mutation score} \(S \;=\; (|M_v| - |M_s|) / |M_v|\), which quantifies how many mutants remain undetected by the current test suite, and 
(2) the structural \emph{coverage} \(C\), calculated as \(C = \frac{|E_{covered}|}{|E_{total}|}\), which captures the ratio of executed structural elements (e.g., lines, branches) to the total number of coverable elements. 

These metrics jointly drive the adversarial interaction between the two agents: 
\begin{itemize}
    \item Agent $\mathcal{T}$ leverages the mutation score \(S\) to refine or regenerate test cases against surviving mutants \(M_s\);
    \item Agent $\mathcal{M}$ leverages both \(S\) and \(C\) to craft new mutants in structurally weak or under-tested regions of the program.
\end{itemize}

Overall, mutation testing acts as the central feedback mechanism in \ourmethod, enabling bidirectional improvement: it strengthens Agent $\mathcal{T}$'s test generation capabilities while guiding Agent $\mathcal{M}$ to synthesize more challenging mutants. This adversarial loop drives the system toward progressively more robust and comprehensive test suites.

\subsection{Test Suite Augmentation}
\label{sec:meth_test_aug}
The mutation testing produces a set of surviving mutants \(M_s\), where each mutant \(m\in M_s\) represents a behavioral variation that the current test suite does not detect. These surviving mutants effectively expose the blind spots of \(T\). 

To fill these blind spots, we augment \(T\) by generating new tests aimed at ``killing'' each surviving mutant.
Specifically, for each \(m\in M_s\), we construct a mutant‐aware prompt that summarizes the mutant in natural language (e.g.\ ``original line: \texttt{return x+y;} mutated to \texttt{return x-y;}''), and provide this prompt to Agent $\mathcal{T}$. The agent is instructed to generate a test case that fails on the mutant variant \(P_m\) while passing on the original program \(P\).

All generated test cases undergo the same test-repair loop as described in Section \ref{sec:meth_initest}. This ensures that only syntax-correct, compilable, and behaviorally valid test cases are included in the augmented suite. Once repaired and validated, the new test is added to \(T\). 
This process is repeated for each surviving mutant, gradually evolving the test suite toward higher fault-detection capability and robustness.

Importantly, we do \emph{not} immediately re-evaluate each new test against its associated mutation after generation. Prior work~\cite{straubinger2025mutation} has shown that while immediate feedback can increase mutant detection rates, it comes at a substantial cost: up to \(\times 7.29\) higher LLM API token usage and significantly increased mutation testing time. 
Instead, we defer evaluation of newly added tests until the next augmentation cycle. Mutants that still survive will be re-targeted in subsequent rounds by Agent \(\mathcal{T}\), allowing us to eventually detect most mutants without incurring excessive computational and financial (the API cost) overhead.

Following this augmentation process, the resulting test suite, $T^+$, is free from syntax and compilation errors. 
This augmented suite is then either passed to the next iteration of the adversarial loop or returned as the final output if termination conditions are met.

\subsection{Mutant Augmentation}
 The agent \(\mathcal{M}\) uses the two key feedback signals obtained from mutation testing: the set of surviving mutants \(M_s\) and the structural coverage map \(C\) to augment its mutant pool in two complementary directions.

\textit{1. Augmentation by Uncovered Code.}
Structural \emph{coverage} \(C\) allows us to extract a set of uncovered lines \(L_u\), which are code regions that remain untested by any case in \(T\). These lines present latent risks, as they may contain faults that the current testing process has yet to examine. Agent \(\mathcal{M}\) proactively attacks these uncovered lines by generating new mutants at those locations, even in the absence of prior mutant feedback. This strategy forces the test suite to interact with previously ignored control paths and execution traces, thereby improving both code coverage and fault exposure.

\textit{2. Augmentation by Surviving Mutants.}
Generating new mutants solely based on structural coverage is often insufficient to provide Agent \(\mathcal{T}\) with actionable feedback. Even when a line of code is covered, it may still conceal faults. For example, when a test case executes a statement without asserting its effects, behavioral deviations remain undetected. As coverage increases, the available space for purely coverage-driven mutation gradually diminishes, limiting the effectiveness of further exploration. 

In this regime, surviving mutants provide a valuable signal. Mutants in \(M_s\) indicate program locations where the current test suite \(T\) fails to detect behavioral divergence from the original program. These locations reveal structural weaknesses in the test suite that are not captured by coverage metrics alone. To exploit this signal, we first group surviving mutants by the specific lines of code they modify, thereby reducing redundancy and focusing mutation efforts on under-tested locations. 
For each group, we construct prompts that instruct the LLM to generate new and diverse mutations on the same line, while varying logic, constants, or operators. This strategy allows Agent \(\mathcal{M}\) to systematically explore a richer space of plausible faults rooted in a shared structural vulnerability, rather than repeatedly mutating already well-tested code.

By integrating mutation generation driven by both surviving mutants and coverage gaps, this augmentation mechanism ensures that the mutant process remains both adaptive (responding to observed weaknesses in the test suite) and exploratory (continuing to probe untested or weakly tested regions). This balance enables a more effective adversarial co-evolution between Agent~\(\mathcal{T}\) and Agent~\(\mathcal{M}\), ultimately leading to more robust and discriminative test suites.

\subsection{Adversarial Iteration Loop}
\label{sec:meth_adviter}
The two agents interact through a structured adversarial loop (Algorithm~\ref{alg:adv-loop}) that runs for a predefined number of rounds $N$.
At each iteration, the agents alternate their actions to progressively improve the test suite quality.

\begin{itemize}
    \item \textbf{Test Suite Augmentation:} Agent $\mathcal{T}$ acts first (Line~5) by generating new tests specifically designed to kill the surviving mutants ($M_s$) identified in the evaluation step.
    \item \textbf{Mutant Augmentation:} Agent $\mathcal{M}$ responds (Line~6) by synthesizing additional mutants to exploit blind spots in the code that remain uncovered or under-tested.
\end{itemize}

Once the loop completes ($i=N$), the final action within the loop corresponds to Agent $\mathcal{M}$. 
To prevent the process from ending with a batch of unchecked mutants, we perform a final round of test suite augmentation (Line~9). 
This ensures that Agent $\mathcal{T}$ always has the final move, guaranteeing that the returned test suite $T$ has responded to the most recent adversarial inputs.

\begin{algorithm}[h]
  \caption{Main adversarial loop of \ourmethod}
  \LinesNumbered
  \small  
  \label{alg:adv-loop}
  \KwIn{Program $P$, prompts $\pi_0,\mu_0,\lambda_0$, max rounds $N$}
  \KwOut{Final test suite $T$}
  
  \SetAlgoLined
  $T \leftarrow \textsc{GenTest}(P,\pi_0,\lambda_0)$ \tcp*{Agent $\mathcal{T}$ plays initial move}
  $M \leftarrow \textsc{GenMutant}(P,\mu_0)$  \tcp*{Agent $\mathcal{M}$ plays initial move}
  
  \For{$i \leftarrow 1$ \KwTo $N$}{
    $(M_s, C, S, M_v) \leftarrow \textsc{MutationTesting}(P,T,M)$ \tcp*{Evaluate the state of the loop}

    \tcp{Agent $\mathcal{T}$ turn: Eliminate surviving mutants}
    $T \leftarrow \textsc{EnhanceTestCaseByMutants}(T, M_s, \pi_0,\lambda_0)$\;
    
    \tcp{Agent $\mathcal{M}$ turn: Exploit blind spots}
    $M \leftarrow \textsc{EnhanceMutantsByFeedback}(M, T, C, \mu_0)$\;
  }
  
  \tcp{Ensure $\mathcal{T}$ responds to $\mathcal{M}$'s last move}
  $(M_s, C, S, M_v) \leftarrow \textsc{MutationTesting}(P,T,M)$\;
  $T \leftarrow \textsc{EnhanceTestCaseByMutants}(T, M_s, \pi_0,\lambda_0)$\;

  \Return $T$\;
\end{algorithm}

\section{Experimental Setup}

We evaluate \ourmethod by addressing the following research questions:
\begin{itemize}
  \item \textbf{RQ1:} How effective is \ourmethod in generating tests in real-world projects?

  \item \textbf{RQ2:} What is the individual contribution of each component of \ourmethod to overall effectiveness?
  \item \textbf{RQ3:} How do iterative rounds affect \ourmethod's effectiveness?
  \item \revtext{\textbf{RQ4:} Does \ourmethod generalize beyond Defects4J to other languages and ecosystems?}

\end{itemize}

\subsection{Datasets}

We evaluate \ourmethod on Defects4J~\cite{defects4j} and GrowingBugs~\cite{GrowingBugsICSE21,GrowingBugsTSE2022,NaturalnessOfBugsFSE2022}, two datasets that provide authentic and reproducible defects in industrial-scale Java code bases. With a total of 20 different projects, 247 different bugs, 727 methods under test. This scale significantly surpasses the evaluations of previous work, such as HITS~\cite{hits} (120 methods) and ChatUniTest~\cite{chatunitest} (264 methods). 

\emph{Defects4J.} Defects4J is a widely adopted benchmark of real, reproducible faults in open source Java projects~\cite{defects4j}. In version 2.1.0, it comprises 835 more bugs drawn from 17 different and high quality projects, each paired with a corresponding fixed version and a comprehensive test suite. Each defect entry includes metadata such as affected files, trigger tests, and developer patches, allowing repeatable fault detection and repair experiments. In this experiment, we used 200 randomly sampled defects from all 17 projects. We also leverage Defects4J framework throughout our experiment and evaluation.

\emph{GrowingBugs.} GrowingBugs is an extensible repository of real faults in open source Java projects built on top of the Defects4J infrastructure~\cite{defects4j}. GrowingBugs automatically filters out non-functional changes from commit histories using the BugBuilder tool, enabling continuous expansion of the dataset without human intervention~\cite{GrowingBugsICSE21,GrowingBugsTSE2022}. Previous studies have shown that patches extracted by the BugBuilder preserve the naturalness of real bugs and support robust empirical evaluations of testing and repair techniques~\cite{NaturalnessOfBugsFSE2022}. To mitigate potential data leakage, we specifically selected all 3 projects that were introduced into the GrowingBugs benchmark after the knowledge cutoff date of the LLM we use for that experiment\revtext{: \texttt{commons-dbcp} (Dbcp), \texttt{commons-beanutils} (Beanutils), and \texttt{commons-fileupload} (Fileupload)}. This subset comprises a total of 47 bugs and 89 methods under test. \revtext{We caution that ``recently added to GrowingBugs'' applies to the benchmark inclusion date, not necessarily to the individual bug introduction dates: some defects in these projects pre-date the LLM training cutoff and could in principle have been observed during pre-training. To further address this residual leakage concern, Section~\ref{sec:rq4} reports an additional generalization study on 12 Python bugs drawn from real GitHub issues post-dating the model cutoff.}

\subsection{Baselines}
We compare \ourmethod against four baselines, two traditional and two LLM-driven: \emph{Randoop}~\cite{Pacheco2007}, a feedback-directed random Java test generator; \emph{EvoSuite}~\cite{evosuite}, a search-based tool that evolves JUnit suites toward high line and branch coverage; \emph{ChatUniTest}~\cite{chatunitest}, an LLM-driven generator that supplies the focal method plus rule-extracted context and uses error-driven test repair (its coverage drops on complex methods~\cite{hits}); and \emph{HITS}~\cite{hits}, an LLM-driven generator that decomposes complex methods into semantically coherent slices and generates tests per slice, achieving the highest LLM-baseline coverage on complex methods. \revtext{We configure EvoSuite and Randoop with a 500 seconds per-bug budget, comparable to \ourmethod's runtime on Defects4J.}

Both LLM baselines use DeepSeek-v3.2 and GPT-OSS-120B \cite{deepseekai2025deepseekv32pushingfrontieropen,openai2025gptoss120bgptoss20bmodel} (we restrict GrowingBugs to DeepSeek-v3, whose cutoff predates the dataset~\cite{hits}). We adapt both for Defects4J: prompts target JUnit 4 instead of JUnit 5, and tests are executed with the Defects4J command. \revtext{Other LLM- and mutation-guided systems are not directly comparable in the context of industrial Java projects. MuTAP~\cite{mutap} targets Python and is evaluated only on student-program benchmarks; Meta’s mutation-guided system~\cite{mutationatmeta} is closed-source; and Alchemist~\cite{barboni2025smartmutant} focuses on Solidity smart contracts.}

\subsection{Metrics}
\label{sec:metrics}
We employ three metrics in our experiments.

\noindent\textbf{(1) Fault Detection Rate} (FDR): 
As our study focuses on the effectiveness of test generation in detecting real faults, the fault detection rate serves as our primary metric~\cite{fdr}. 
Let $\mathcal{F} = \{f_1, f_2, \ldots, f_N\}$ be a set of $N$ known defects, where each $f_i$ has a fixed (reference) version $P_i^{\mathit{fix}}$ and a buggy version $P_i^{\mathit{bug}}$, let $T_i$ be the test suite generated for $f_i$, and let $\text{Fail}(T, P)$ denote the tests in $T$ that fail on program $P$. The detection indicator is defined as:  
  \[
  \text{Detect}(T_i, f_i) =
  \begin{cases}
    1, & \text{if } \text{Fail}(T_i, P_i^{\mathit{bug}}) \setminus \text{Fail}(T_i, P_i^{\mathit{fix}}) \neq \varnothing \\
    0, & \text{otherwise}
  \end{cases}
  \]
Then, FDR is computed as:
\[
\text{FDR} = \frac{1}{N} \sum_{i=1}^{N} \text{Detect}(T_i,f_i)
\]
\newrev{All methods, including the baselines, generate $T_i$ on the fixed version $P_i^{\mathit{fix}}$; the buggy version $P_i^{\mathit{bug}}$ is used only to score detection by replaying the suite. Section~\ref{sec:threats} discusses this reference-version setup.}

\noindent\textbf{(2) Coverage}: In addition to fault detection, we also report the coverage of each generated suite (measured with Cobertura\cite{cobertura-tool}) in accordance with previous works~\cite{chatunitest,hits}. We report both \emph{line} and \emph{branch coverage}. \emph{Line coverage} measures the percentage of lines that have been executed by the test cases. \emph{Branch coverage} measures the percentage of branches (decision points) in the source code that have been executed during the testing process~\cite{jorgensen2013software}.

\noindent\textbf{(3) Cost}: We also evaluate the API token cost of LLM-based methods. For API access, we use the official DeepSeek platform for DeepSeek models and Tinker for GPT-OSS models. We track token usage throughout the entire generation process of each method and report the average cost per method on each dataset.

\subsection{Parameter Configuration}
We run our method on multiple backbone LLMs, including DeepSeek-v3.2, GPT-OSS-120B for both agents and LLM-based baselines. We choose these models for their balance between strong coding ability and low cost. To minimize randomness in test generation, we set \texttt{temperature=0} as suggested for all LLM-based test generation methods \cite{liu2024deepseek}. 
All tools and generated tests are built and executed under Java 8 with JUnit 4 and Mockito 4.11 to ensure compatibility with Defects4J.  
We limit the adversarial loop to a maximum of 5 iterations. Excluding the initial generation process, this comprises a combined total of 5 actions between the agents, beginning and ending with Agent \(\mathcal{T}\). \revtext{Agent~$\mathcal{M}$'s per-invocation mutant budget is set adaptively to $\max(2\,\textit{line\_num},\,50)$, scaling with method length.}
To measure coverage, all suites are instrumented and measured with Cobertura \cite{cobertura-tool} using its default configuration provided by Defects4J framework\cite{defects4j}.

\section{Results and Analysis}

\begin{table*}[t]
\caption{Comparison of Fault Detection Rate (FDR), Coverage, and Cost on Defects4J and GrowingBugs Datasets. Best results for each model are highlighted in \textbf{bold}. Second best results are \underline{underlined}.}
\label{tab:combined_results}
\centering
\small
\resizebox{\textwidth}{!}{
\begin{tabular}{l cccc cccc}
\toprule
 \multirow{2}{*}{\textbf{Method}}& \multicolumn{4}{c}{\textbf{Defects4J}} & \multicolumn{4}{c}{\textbf{GrowingBugs}} \\
\cmidrule(lr){2-5} \cmidrule(lr){6-9}
 & \textbf{FDR} & \textbf{Line Cov.} & \textbf{Branch Cov.} & \textbf{Cost} & \textbf{FDR} & \textbf{Line Cov.} & \textbf{Branch Cov.} & \textbf{Cost} \\
\midrule
\multicolumn{9}{c}{\cellcolor{lb}\textit{Traditional Approaches}} \\
\textbf{Randoop}   & \revtext{30.21\%} & \revtext{38.39\%} & \revtext{28.99\%} & -- & \revtext{23.40\%} & \revtext{38.16\%} & \revtext{27.34\%} & -- \\
\textbf{EvoSuite}  & \revtext{44.36\%} & \revtext{69.45\%} & \revtext{61.54\%} & -- & \revtext{40.43\%} & \revtext{67.86\%} & \revtext{64.37\%} & -- \\

\multicolumn{9}{c}{\cellcolor{lb}\textit{LLM-Based Approaches}} \\
\textbf{GPT-OSS-120B} & & & & & & & & \\
\hspace{3mm} ChatUniTest & 21.56\% & 18.89\% & 17.83\% & \textbf{\$0.178} & -- & -- & -- & -- \\
\hspace{3mm} HITS     & 33.93\% & 28.33\% & 25.97\% & \$1.096 & -- & -- & -- & -- \\
\hspace{3mm} \textbf{AdverTest (Ours)}     & \bf 57.05\% & \bf 60.50\% & \textbf{57.40\%} & \underline{\$0.553} & -- & -- & -- & -- \\

\addlinespace[4pt]
\textbf{DeepSeek} & & & & & & & & \\
\hspace{3mm} ChatUniTest & 13.52\% & 14.38\% & 14.15\% & \textbf{\$0.113} & 31.91\% & 32.07\% & 34.79\% & \textbf{\$0.089} \\

\hspace{3mm} HITS & \underline{61.38\%} & \underline{60.13\%} & \underline{52.40\%} & \$0.411 & \underline{61.70\%} & \textbf{77.99\%} & \underline{66.02\%} & \$0.407 \\
\hspace{3mm}  \textbf{AdverTest (Ours)}  & \textbf{66.63\%} & \textbf{62.26\%} & \bf 57.06\% & \underline{\$0.270} & \textbf{65.96\%} & \underline{74.54\%} & \textbf{70.53\%} & \underline{\$0.245} \\

\bottomrule
\end{tabular}
}
\end{table*}

\subsection{RQ1: Effectiveness}
\label{sec:rq1}
We evaluate the effectiveness of \ourmethod by comparing its FDR and coverage against traditional and LLM-based baselines. The results are summarized in Table~\ref{tab:combined_results}.
\paragraph{Fault Detection Capability}
\ourmethod consistently achieves high fault detection rates across all datasets. On the Defects4J dataset, \ourmethod (using DeepSeek V3.2) attains an FDR of \textbf{66.63\%}. This represents a relative improvement of approximately \textbf{8.6\%} over the state-of-the-art LLM-based method, HITS (61.38\%), and a substantial \revtext{\textbf{50.20\%}} improvement over the traditional search-based tool, EvoSuite \revtext{(44.36\%)}.
A similar trend is observed on the GrowingBugs dataset, where \ourmethod achieves \textbf{65.96\%} FDR compared to 61.70\% for HITS and \revtext{40.43\%} for EvoSuite. This result is particularly significant, as GrowingBugs contains defects introduced more recently, serving as a rigorous test for data leakage and overfitting.
\paragraph{Coverage}
\revtext{In terms of coverage, EvoSuite achieves the highest line and branch coverage on Defects4J under the matched 500\,s budget, with \ourmethod a close second on both metrics. This higher coverage, however, does not translate into higher fault detection: EvoSuite's FDR (44.36\%) still trails \ourmethod's 66.63\% by a wide margin.} EvoSuite primarily optimizes for code execution (coverage and weak mutation), often generating tests that reach a faulty line without propagating the error to an observable output~\cite{evosuite}.
In contrast, \ourmethod integrates strong mutation directly into the loop: surviving mutants explicitly guide the LLM to generate tests that distinguish the mutant's behavior from the original code, ensuring the generated tests not only execute the code but also rigorously expose semantic faults.

\revtext{On GrowingBugs, HITS leads line coverage thanks to its slicing-based method, while \ourmethod leads branch coverage and FDR.} Unlike HITS~\cite{hits}, \ourmethod does not focus solely on high-coverage test cases.

\paragraph{Robustness Across Foundation Models}
As shown in Table~\ref{tab:combined_results}, \ourmethod demonstrates superior robustness across both LLMs. Even with the weaker GPT-OSS-120B model, \ourmethod maintains a high FDR of \textbf{57.05\%} whereas the FDR of HITS drops from 61.38\% to 33.93\%. This suggests that the adversarial feedback loop effectively compensates for the weaker reasoning capabilities of smaller models, whereas HITS's slicing method relies heavily on the raw capability of the foundation model.

\paragraph{Cost analysis}
Among LLM baselines (Table~\ref{tab:combined_results}, cost columns), ChatUniTest incurs the lowest absolute API cost but at the expense of test quality. Against the strongest baseline HITS, \ourmethod reduces per-method API cost by \textbf{34.3\%} on Defects4J w/ DeepSeek-v3.2 (\$0.270 vs.\ \$0.411), \textbf{49.5\%} on Defects4J w/ GPT-OSS-120B (\$0.553 vs.\ \$1.096), and \textbf{39.8\%} on GrowingBugs (\$0.245 vs.\ \$0.407). Absolute prices depend on platform pricing policies, but the relative advantage holds across all settings.

\revtext{\paragraph{Runtime} Per-method wall-clock time on Defects4J: EvoSuite and Randoop use the matched 500\,s budget; among LLM-based methods, ChatUniTest averages 122.7\,s, \ourmethod 484.95\,s, and HITS 878.63\,s. \ourmethod is therefore roughly $1.8\times$ faster than HITS while detecting more bugs, and runs within the same time envelope as the search-based baselines.}

\paragraph{Statistical Significance}
A McNemar's test (appropriate for matched-paired binary detection outcomes) against the strongest baseline HITS, combining both DeepSeek V3.2 and GPT-OSS-120B results on Defects4J, yields \textbf{95} bugs that \ourmethod detects but HITS misses versus \textbf{57} the other way, with $p=\textbf{0.00257}$. We therefore reject the null hypothesis at $p<0.01$ and conclude that the FDR gain is statistically significant.

\revtext{\paragraph{Mutant Quality} On Defects4J, Agent~$\mathcal{M}$'s output achieves a \textbf{90.3\%} compile rate, \textbf{36.2\%} survival rate among valid mutants, and an equivalent-mutant rate of \textbf{3.45\%}---comparable to the \textbf{2.1\%} reported for the MAJOR mutation tool~\cite{WangMutation2024}.}

\begin{tcolorbox}[enhanced, width=\linewidth, boxrule=0.8pt, left=2pt, right=2pt, top=2pt, bottom=2pt,]
\textbf{Answer to RQ1.} \ourmethod significantly outperforms both search-based and LLM-based baselines, improving fault detection rates by up to \revtext{50.20\%} over EvoSuite and 8.6\% over HITS on Defects4J, with consistent generalizability on GrowingBugs. 
Our method also proves cost-effectiveness by reducing API costs by more than 34.3\% compared to HITS. Furthermore, the framework proves highly robust, leveraging the adversarial loop to maintain high effectiveness even when utilizing less capable foundation models.
\end{tcolorbox}

\subsection{RQ2: Ablation Study}
\label{sec:res:rq2}
To isolate the contribution of each core component in our framework, we conduct an ablation study on a subset of Defects4J dataset with GPT-OSS-120B as the base LLM. This subset contains a total of 50 \revtext{bugs randomly sub-sampled from the same 200-bug Defects4J set used in RQ1, comprising} 129 methods under test from all 17 Defects4J projects. \revtext{The sampling is chosen to keep the per-variant compute cost tractable; we discuss the resulting representativeness threat in Section~\ref{sec:threats}.} Specifically, we evaluate the impact of \revtext{four} critical components:

(1) Adversarial Iterative Loop: We disable the iterative co-evolution process between Agent \(\mathcal{T}\) and \(\mathcal{M}\), limiting the system to a single round of initial LLM-based test generation without any adversarial feedback or augmentation; and

(2) Surviving Mutant Feedback: We remove the fine-grained feedback mechanism introduced in Section \ref{sec:meth_test_aug}, which informs the test generator of the exact nature of surviving mutants. In this setting, the LLM is aware that certain mutants remain undetected but receives no details about their specific locations or semantics. It must therefore generate additional tests without targeted guidance.

\revtext{(3) Coverage-only Feedback: We replace all mutation-based signals with coverage information alone. Agent~$\mathcal{T}$ receives the uncovered source lines (with their code) and the affected method bodies in place of any surviving-mutant descriptions. This variant isolates the contribution of mutation-based feedback over a pure coverage-driven signal.}

\revtext{(4) Classical Mutation Substitute: We replace Agent~$\mathcal{M}$ with the MAJOR mutation tool~\cite{Just2011} while leaving every other component of the loop unchanged. Only the source of mutants changes from LLM-generated to rule-based; the feedback channel, the iteration schedule, and Agent~$\mathcal{T}$ are identical to the full pipeline. This variant isolates the contribution of LLM-driven mutation over a classical rule-based mutation engine.}

\begin{table}[h]
  \centering
  \caption{Ablation Study on the Defects4J Dataset}
\label{tab:ablation-math}
  \small
  \setlength{\tabcolsep}{10pt}
    \begin{tabular}{lccc}
      \toprule
         \multirow{2}{*}{Variant methods}&\multirow{2}{*}{FDR} & \multicolumn{2}{c}{Coverage} \\
       \cline{3-4}
        &  & Line & Branch \\
      \midrule
      \ourmethod       & \textbf{54.00}    & \textbf{88.17}         & \textbf{82.01}          \\
      w/o Iter    & 26.00             & 68.62         & 56.79          \\
      w/o Mut     & 40.00             & 77.52           & 66.22  \\
      \revtext{w/ Cov-only} & \revtext{34.00}      & \revtext{80.33}        & \revtext{69.72}         \\
      \revtext{w/ MAJOR}    & \revtext{46.00}      & \revtext{85.71}        & \revtext{80.12}         \\
      \bottomrule
    \end{tabular}%
\end{table}
Table~\ref{tab:ablation-math} shows that removing the iterative loop (\textbf{w/o Iter}) incurs the largest drop in FDR (51.85\%) and coverage (22.17\% line / 30.75\% branch). Omitting mutant‑aware prompting (\textbf{w/o Mut}) also degrades performance substantially, confirming that providing the LLM with concrete mutation details is critical for generating effective tests. However, 'w/o Mut' still achieves higher metrics compared to 'w/o Iter', that is because we still provide Agent $\mathcal{T}$ chances to improve the test case, even without specific information, knowing surviving mutants exists is a benefit. Also, missing mutant information will lower the mutation score and the ability to 'kill' surviving mutants, which will result in more test case enhancement rounds.

\revtext{Replacing mutation feedback with coverage-only signals (\textbf{w/ Cov-only}) drops FDR to 34.00\%, which is lower than \textbf{w/o Mut}'s 40.00\% despite slightly higher coverage (80.33\% line). This suggests that surviving-mutant feedback carries information beyond what coverage signals alone provide. Substituting the LLM mutator with the rule-based MAJOR tool (\textbf{w/ MAJOR}) loses 8\,pp FDR (from 54.00\% to 46.00\%) while coverage remains nearly intact (85.71\% line), suggesting that the LLM mutator's contribution stems more from qualitative mutant diversity than from reaching new lines.}

The results indicate that all four ablation variants degrade performance: \revtext{ablating or substituting any single component reduces FDR by 14.81\%--51.85\% and coverage by 2.30\%--30.75\%.}

\begin{tcolorbox}[enhanced, width=\linewidth, boxrule=0.8pt, left=2pt, right=2pt, top=2pt, bottom=2pt,]
\textbf{Answer to RQ2.} \revtext{All four ablation variants confirm \ourmethod's design: the iterative loop is the single largest driver of FDR; mutation-based feedback contributes information that pure coverage signals cannot capture; and LLM-driven mutation outperforms a classical rule-based mutator while keeping coverage comparable.}
\end{tcolorbox}

\subsection{RQ3: Effect of Iterative Rounds}
\begin{figure}[t] 
\centering
    \includegraphics[width=0.9\linewidth, trim=0 20 0 20 clip]{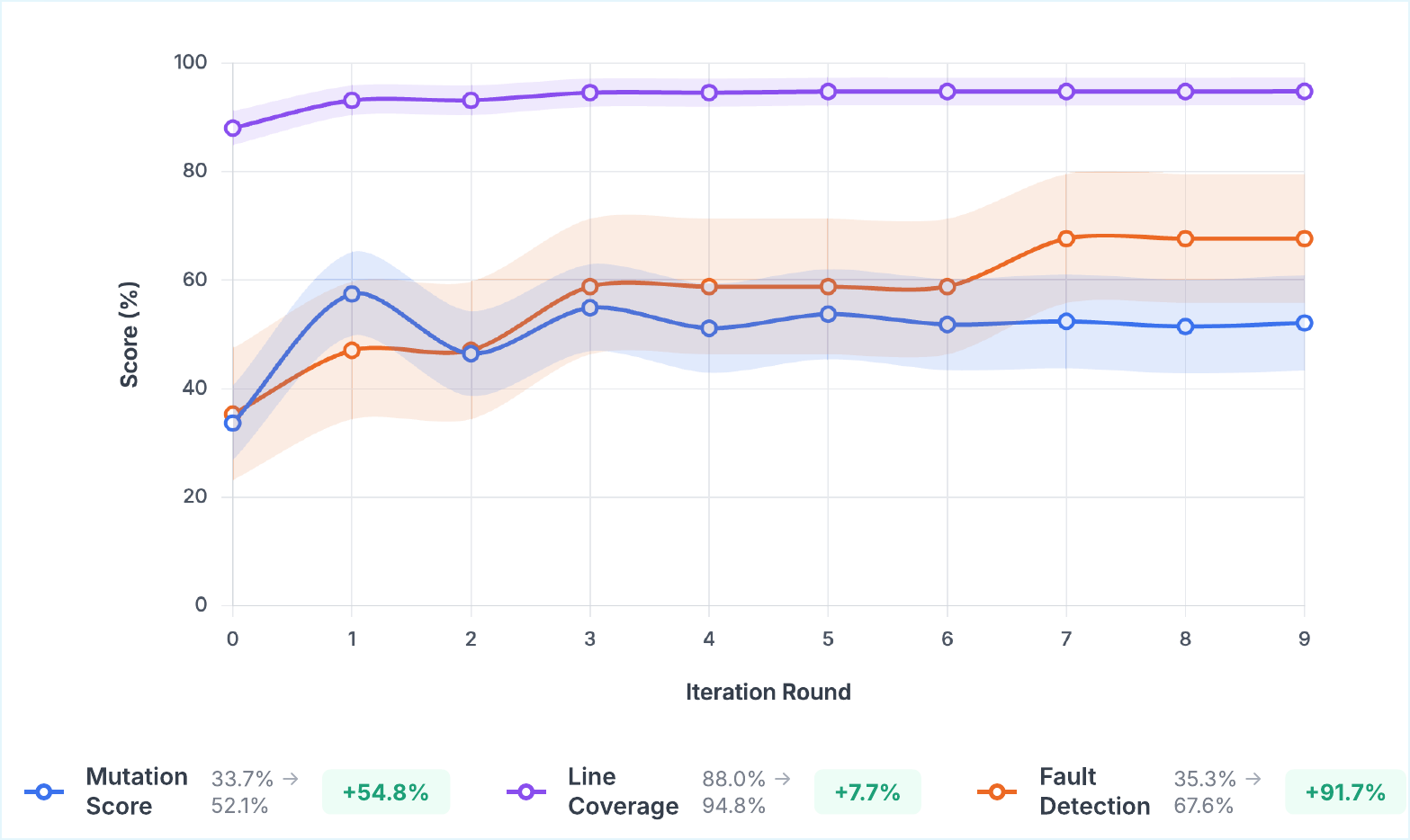} 
    \caption{Mutation Score (MS), Line Coverage (CV), and Fault Detection Rate across Nine Rounds. The shadow indicates the standard deviation.}
    \label{fig:iter_rounds} 
\end{figure}

To analyze the impact of iteration rounds on \ourmethod's performance, we executed the adversarial loop for a total of 9 rounds on the subset defined in Section \ref{sec:res:rq2}, utilizing GPT-OSS-120B as the underlying model. In this setup, a single ``round'' corresponds to one action by either Agent $\mathcal{T}$ or Agent $\mathcal{M}$. 

Figure~\ref{fig:iter_rounds} illustrates the co-evolution of the metrics. Round 0 represents the performance of the initial test suite generated. In subsequent steps, the agents alternate: Agent $\mathcal{T}$ performs test augmentation in odd rounds (1, 3, 5, 7, 9), while Agent $\mathcal{M}$ generates supplementary mutants in even rounds (2, 4, 6, 8).

We observe distinct behaviors across the three metrics:

\textbf{Line Coverage (Purple):} Coverage shows a steady but modest increase, rising from 88.0\% to 94.8\% (+7.7\%). The high initial starting point suggests that modern LLMs are inherently proficient in achieving structural coverage. Consequently, the marginal gains diminish in later rounds as the code becomes saturated.

\textbf{Mutation Score (Blue):} The Mutation Score (MS) displays a sawtooth pattern characteristic of the adversarial process. During Agent $\mathcal{T}$'s turns (odd rounds), MS increases as the test suite is refined to kill existing mutants. Conversely, during Agent $\mathcal{M}$'s turns (even rounds), MS decreases as new mutants are injected to exploit blind spots. Despite these local fluctuations, the global trend is a significant net increase of 54.8\%, indicating that the test suite is becoming progressively more robust against semantic faults.

\textbf{Fault Detection Rate (Orange):} The most substantial improvement is observed in the FDR, which surges from 35.3\% to 67.6\% (+91.7\%). Notably, while coverage increases moderately, FDR continues to rise significantly in later iterations (e.g., the jump at Round 3 and 7). This confirms that the mutation guided adversarial loop successfully directs the agents toward corner cases and logical faults that coverage metrics overlook and generate more robust test cases.

\begin{tcolorbox}[enhanced, width=\linewidth, boxrule=0.8pt, left=2pt, right=2pt, top=2pt, bottom=2pt,]
\textbf{Answer to RQ3.} \ourmethod's effectiveness improves markedly with iterative rounds. While coverage gains are incremental, the adversarial interaction drives a substantial increase in Mutation Score and Fault Detection Rate, demonstrating that the iterative loop is essential for detecting complex, real-world faults.
\end{tcolorbox}

\subsection{\revtext{RQ4: Generalization Beyond Defects4J}}
\label{sec:rq4}

\revtext{To verify that the adversarial loop transfers beyond Java and Defects4J, we re-implemented \ourmethod for Python and evaluated it on the SWE-Rebench benchmark~\cite{badertdinov2026swe}, which is continuously updated from live GitHub issues. We used 12 instances from the SWE-Rebench 2026-02 leaderboard, all of which post-date the LLM training cutoff, providing a stronger data-leakage control than relying on benchmark inclusion dates alone (cf.\ Section~\ref{sec:threats}).}

\paragraph{\revtext{Setup}}
\revtext{The 12 instances span AWS SDKs (\texttt{bedrock-agentcore-sdk-python}, \texttt{moto}), agent frameworks (\texttt{a2a-python}), build systems (\texttt{conan}), NLP pipelines (\texttt{haystack}), document parsing (\texttt{docling}), filesystem mocking (\texttt{pyfakefs}), ORM tooling (\texttt{advanced-alchemy}, \texttt{pymilvus}), Django templating (\texttt{django-bird}), deep-learning libraries (\texttt{keras}), and notebook runtimes (\texttt{marimo}). We ported the full pipeline of Sections~\ref{sec:meth_initest}--\ref{sec:meth_adviter} to \texttt{pytest} and \texttt{coverage.py}, keeping all algorithmic components. The backbone LLM is GPT-OSS-120B. Detection follows the same criterion as in Section~\ref{sec:metrics}: the generated suite \emph{detects} the bug iff some test fails on the buggy version but not on the fixed version.}

\paragraph{\revtext{An Additional Mutant Source}}
\revtext{In porting \ourmethod to Python, we observed a recurring failure mode of the two-source mutant augmentation described in Section~\ref{sec:meth_test_aug}. \newrev{Many covered, bug-relevant lines had no mutant attached, because the initial call concentrated mutants on a few operators per line, so $\textsc{EnhanceTestCaseByMutants}$ never received a survivor at those locations.} We therefore enabled a third augmentation source for the Python pipeline. In addition to (S1)~variations of surviving mutants and (S2)~mutants on uncovered lines, the agent now also generates (S3)~mutants on lines that are covered by the current suite but have no mutant associated with them. We treat S3 as an opt-in extension of the methodology and report both pipeline variants below.}

\paragraph{\revtext{Results}}
\revtext{Table~\ref{tab:rq4_python} reports the head-to-head comparison. Enabling the untouched-line source raises FDR from 7/12 (\textbf{58.3\%}) to \textbf{11/12} (\textbf{91.7\%}); \newrev{the four additional bugs become detectable once the round-3 $\textsc{EnhanceTestCaseByMutants}$ step receives survivors generated by S3.} Across the 11 detected instances, 6/11 (54.5\%) require at least one enhancement round to surface, and 2/11 (18.2\%) are detected only at round~3 or later, where the alternation between S3-driven mutant injection and Agent~$\mathcal{T}$'s test refinement is essential.}

\begin{table}[t]
\caption{\revtext{Generalization on 12 SWE-Rebench Python instances (GPT-OSS-120B; MS and coverage averaged across instances for the full pipeline). S1: variations of surviving mutants. S2: mutants on uncovered lines. S3: mutants on covered-but-untouched lines.}}
\label{tab:rq4_python}
\centering
\small
\setlength{\tabcolsep}{6pt}
\begin{tabular}{lcccc}
\toprule
\textbf{Pipeline} & \textbf{Detected} & \textbf{FDR} & \textbf{Avg MS} & \textbf{Avg Cov.} \\
\midrule
\revtext{S1 + S2}        & \revtext{7/12}            & \revtext{58.3\%}          & \revtext{---}          & \revtext{---}        \\
\revtext{S1 + S2 + S3}   & \revtext{\textbf{11/12}}  & \revtext{\textbf{91.7\%}} & \revtext{64.2\%}       & \revtext{93.0\%}     \\
\bottomrule
\end{tabular}
\end{table}

\paragraph{\revtext{Why S3 Helps}}
\newrev{S3 attaches mutants to covered lines that carry no operator-style mutant, surfacing survivors that drive Agent~$\mathcal{T}$ to probe their semantics. For example, in \texttt{deepset-ai/haystack\#10547}, mutating an \texttt{except TypeError} clause to \texttt{except BaseException} forced Agent~$\mathcal{T}$ to write a test that probes the same semantic boundary as the real defect. We note that this hint is available only because generation operates on the fixed version, which retains the \texttt{except} clause that the buggy code omits; we discuss this reference-version setup and its implications in Section~\ref{sec:threats}. The one undetected instance (\texttt{getmoto/moto\#9727}) reached \textbf{83.8\%} coverage yet went undetected, a reminder that high coverage alone does not guarantee fault detection.}

\paragraph{\revtext{Leakage Controls}}
\revtext{Throughout the loop, the LLM receives only the source code of the \emph{fixed} target function, the file path, and module-level context. It never sees the GitHub issue description, the developer patch, the developer-written tests, the hint text, or the \texttt{FAIL\_TO\_PASS} oracle from SWE-Rebench. The only bug-specific information used by the pipeline is the function-level scope (which method the patch touches), which is the same level of supervision \ourmethod uses on Defects4J. The \textbf{91.7\%} FDR is therefore obtained on bugs whose patches were not in the model's training data and whose oracles the pipeline does not see.}

\begin{tcolorbox}[enhanced, width=\linewidth, boxrule=0.8pt, left=2pt, right=2pt, top=2pt, bottom=2pt]
\revtext{\textbf{Answer to RQ4.} \ourmethod transfers from Java/Defects4J to Python/SWE-Rebench, detecting \textbf{11/12} (\textbf{91.7\%}) of real-world Python bugs drawn from post-cutoff GitHub issues. This suggests that the gains are not narrowly tied to language, test-tooling, or model memorization.}
\end{tcolorbox}

\subsection{Case Study}
\begin{figure*}[t]
  \centering
  \includegraphics[width=\linewidth]{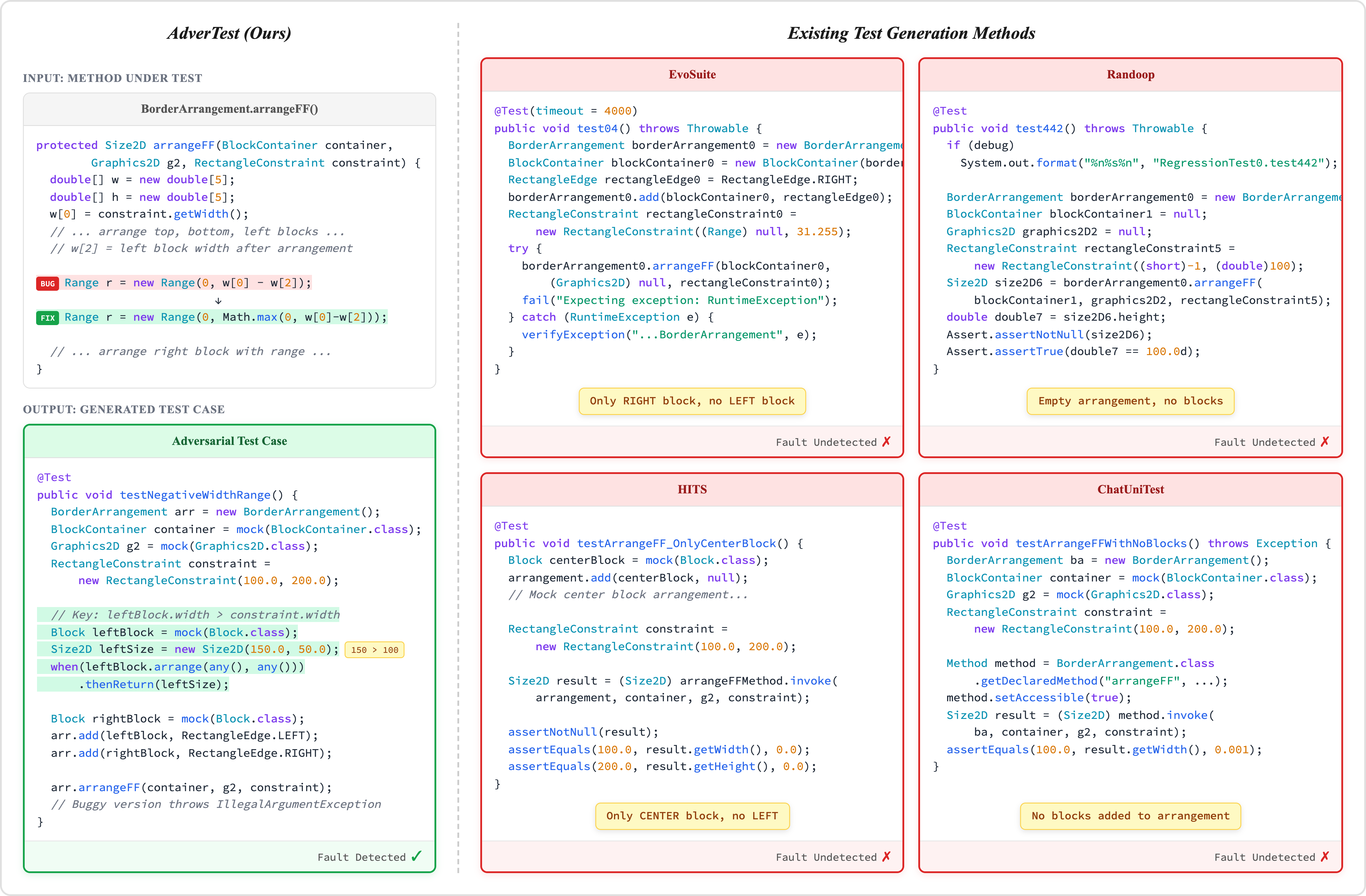}
  \caption{An Example of Fault Detection Process for \texttt{arrangeFF}.}
  \Description{Figure for case study}
  \label{fig:arrangeFF}
\end{figure*}

To demonstrate the effectiveness of \ourmethod, we present a representative case study involving a defect in \texttt{jfree-chart} project from \textbf{Defects4J} dataset. We selected this case because the fault involves a boundary condition easily overlooked by standard coverage metrics, yet it \newrev{illustrates how the components of \ourmethod work together}. As illustrated in Figure~\ref{fig:arrangeFF}, \newrev{among the evaluated tools only \ourmethod detected this bug}.

The top-left panel of Figure~\ref{fig:arrangeFF} displays the method under test, \texttt{arrangeFF}, along with the fix. This method arranges blocks within a container subject to fixed width and height constraints. The buggy version fails to account for the scenario where the width of the left block exceeds the total constraint width, producing a negative upper bound for the right block's range (calculated as \texttt{w[0] - w[2]}) that triggers an \texttt{IllegalArgumentException}. This bug is subtle because standard inputs easily cover the faulty statement without triggering the exception, masking the untested critical boundary condition.

At first, Agent $\mathcal{T}$'s initial test generation failed to detect the bug, as the input space was too broad for the LLM to effectively target. Similarly, all baseline methods failed; as shown in figure\ref{fig:arrangeFF}, tools like EvoSuite, Randoop, HITS, and ChatUniTest generated various block layouts, but none produced the specific condition where the left block is wider than the container. However, Agent $\mathcal{M}$ successfully generated 70 mutants, two of which effectively simulated the underlying defect (e.g., by replacing \texttt{Math.max} with \texttt{Math.min} or removing the function call entirely). These mutants survived the initial test suite. \newrev{Recall that \ourmethod generates on the fixed version, so Agent~$\mathcal{M}$ produces these mutants by perturbing the corrected code, recreating the historical fault.}

In the subsequent test augmentation phase, these surviving mutants provided critical guidance to Agent $\mathcal{T}$. By analyzing the surviving mutants, Agent $\mathcal{T}$ inferred the necessity of a test case where the constraint width is smaller than the left block's width (\texttt{w[2]}). Consequently, as shown in the bottom-left panel, Agent $\mathcal{T}$ generated a targeted adversarial test case that satisfied this condition, successfully exposing the fault.

\newrev{This example illustrates how mutation guidance and the LLM's semantic understanding work together.} First, unlike EvoSuite and Randoop, which produce tests with low readability and maintainability~\cite{deljouyi2024leveraging}, \ourmethod generates semantically meaningful and readable test code. More importantly, the surviving mutants effectively directed Agent $\mathcal{T}$ to focus on the precise input region required to trigger the fault. Without such guidance, the test search space remains too broad, making the probability of randomly generating a fault-revealing input negligible. Furthermore, this highlights the advantage of LLM-based mutation: unlike traditional operators, LLMs leverage code context to produce compilable, logic-altering mutations—such as changing \texttt{Math.max} to \texttt{Math.min}—that drive deeper testing.

\section{Threats to Validity}
\label{sec:threats}

We organize threats to validity following established categories in empirical software engineering.

\textbf{Construct Validity.}  
We measure effectiveness using fault detection rate and coverage metrics (line and branch). Fault detection rate relies on the accuracy of defect annotations in Defects4J~\cite{defects4j} and GrowingBugs~\cite{GrowingBugsICSE21,GrowingBugsTSE2022,NaturalnessOfBugsFSE2022}; any missing or mislabeled defects may bias results. Coverage metrics depend on the instrumentation tool (Cobertura~\cite{cobertura-tool}); failures to instrument generated tests or exclusions in configuration could lead to under- or over-reporting.

\newrev{\textbf{Reference-Version Test Generation.} All methods, including the baselines, generate tests (and, for \ourmethod, mutants) on the \emph{fixed} (reference) version of each program, and detection is scored by replaying the suite on the buggy version (Section~\ref{sec:metrics}). Generating every tool against a shared reference is the standard, fair basis for comparison, since a tool given the buggy version would simply encode the buggy behavior as expected. A consequence is that a mutant can occasionally land on a line introduced by the fix and entirely absent from the buggy version, as in the \texttt{try/except} block of the SWE-Rebench example in Section~\ref{sec:rq4}, where the hint is available only because generation occurs on the fixed version. We regard this as a genuine limitation rather than a confound: for most defects no mutant lands on fix-introduced code, and mutation instead helps through the coupling effect, whereby a suite that kills the seeded mutants is sensitive enough to expose the real fault. This reflects the premise of mutation testing, on which our framework is built: faults are detected as deviations from a program taken to be correct, with the real buggy version playing the same role that seeded mutants play during generation.}

\textbf{Internal Validity.}
We mitigate threats to internal validity through several measures. First, \ourmethod is compared against four state-of-the-art baselines~\cite{Pacheco2007,evosuite,hits,chatunitest}, and statistical analysis is applied to ensure significance. Experiments are replicated across multiple LLMs to verify that observed gains are not artifacts of a specific model.
Second, we address potential \textit{data leakage}. Defects were intentionally selected from GrowingBugs entries added in December 2024, after the DeepSeek V3~\cite{liu2024deepseek} knowledge cutoff, to reduce the risk of models having prior exposure. Nevertheless, parts of the underlying projects may have existed earlier and could have been seen by LLMs, which may still introduce subtle leakage effects.
\revtext{Third, the RQ2 ablation (Section~\ref{sec:res:rq2}) uses a 50-bug random sub-sample of RQ1's 200 bugs for compute reasons; we therefore read it only for component-wise comparisons under matched conditions, not as absolute effectiveness claims.}
Finally, while LLM nondeterminism presents an inherent validity threat, our diverse model selection and large-scale testing help ensure the robustness of reported results.

\textbf{External Validity.}
\revtext{Our evaluation spans both Java and Python, covering Defects4J~\cite{defects4j} and GrowingBugs~\cite{GrowingBugsICSE21,GrowingBugsTSE2022,NaturalnessOfBugsFSE2022} for Java as well as 12 post-cutoff real-world issues from SWE-Rebench~\cite{badertdinov2026swe} for Python. These results provide initial evidence that \ourmethod is not tied to a single language, benchmark, or test framework. Nevertheless, the current study still covers only two programming ecosystems and a limited set of tools (e.g., JUnit/Cobertura~\cite{cobertura-tool} and pytest/coverage.py); the Python generalization study is also smaller in scale than the Java evaluation. Regarding model selection, we evaluate \ourmethod on three representative LLMs to approximate broader applicability, but computational constraints prevent exhaustive evaluation of the entire model landscape. Caution is therefore still needed when extrapolating our findings to other languages, domains, testing frameworks, mutation engines, or future LLM architectures.}

\section{Future Work}
\paragraph{Extending to Higher-Order Mutants}
While our current framework focuses on first-order mutants for diagnosability and prompt reliability, a natural evolution is the generation of Higher-Order Mutants (HOMs)~\cite{jia2009higher}, either by combining LLM-generated first-order mutants or by prompting the LLM for multi-line edits directly. HOMs can simulate subtle faults that arise from the interaction of multiple defects, but the added complexity may not be cost-effective: the Coupling Effect hypothesis~\cite{OFFUT92COUPLING} suggests that test suites strong enough to kill simple mutants are already likely to kill complex ones. We leave a systematic study of this trade-off to future work.

\revtext{\paragraph{RIP-Aware Survivor Feedback.} Currently, when a mutant survives, we forward only its source-level diff to Agent $\mathcal{T}$ and let the LLM infer \emph{why} the existing suite failed to kill it. A richer feedback channel would explicitly distinguish the three stages of the Reachability-Infection-Propagation (RIP) model: a survivor may go unreached (no test executes the mutated line), uninfected (executed but produces the same intermediate state), or unpropagated (state diverges but the divergence never reaches an observable assertion). Automating this RIP classification with lightweight dynamic analysis and presenting it alongside the mutant would let Agent $\mathcal{T}$ pick targeted remediations (extend coverage, strengthen assertions, or probe new outputs) rather than guessing. Dynamically computing the precise reason a mutant survives is a non-trivial program-analysis problem; we leave a thorough study of cost-effective RIP feedback to future work.}

\revtext{\paragraph{Beyond Unit-Test Generation.} The core philosophy of \ourmethod---two LLM agents iteratively probing each other's blind spots under structural and mutation feedback---is, in principle, not tied to unit testing. A natural extension is compiler fuzzing (e.g., WhiteFox~\cite{yang2024whitefox}), where the connection between a mutation in the compiler source and the input program that distinguishes it is far less direct than a method call against a focal class. Adapting the feedback loop to such domains would require richer signals than the source diff and coverage map we use today---likely including IR-level execution traces and differential output checks---but we see this as a promising avenue for generalizing the adversarial dual-agent paradigm to more complex testing tasks.}

\section{Conclusion}
In this paper, we have presented \ourmethod, an adversarial dual-agent framework to generate high-quality, robust unit tests. \ourmethod combines a test generation agent ($\mathcal{T}$) with a mutant generation agent ($\mathcal{M}$), guiding their interaction through bidirectional feedback on mutation score and line coverage. The adversarial loop systematically exposes blind spots in the evolving test suite and drives both agents toward stronger fault detection capability.
Experiments on two real-world benchmarks, Defects4J and GrowingBugs, demonstrate the practical benefits of this design.  
\ourmethod improves the fault detection rate with statistical significance by 8.56\% over the best existing LLM-based approach HITS and by \revtext{50.20\%} over the search-based tool EvoSuite, while maintaining a very competitive coverage rate against the state-of-the-art method HITS.

\section*{Data Availability}

All code and data used in this study are publicly available at \url{https://github.com/jmueducn/AdverTest}. 

\section*{Acknowledgment}
This research is funded by the National Key Research and Development Program of China (Grant No. 2023YFB4503802) and the Natural Science Foundation of Shanghai (Grant No. 25ZR1401175).

\bibliographystyle{ACM-Reference-Format}
\bibliography{main}

@inproceedings{evosuite,
author = {Fraser, Gordon and Arcuri, Andrea},
title = {EvoSuite: automatic test suite generation for object-oriented software},
year = {2011},
isbn = {9781450304436},
publisher = {Association for Computing Machinery},
address = {New York, NY, USA},
url = {https://doi.org/10.1145/2025113.2025179},
doi = {10.1145/2025113.2025179},
booktitle = {Proceedings of the 19th ACM SIGSOFT Symposium and the 13th European Conference on Foundations of Software Engineering},
pages = {416–419},
numpages = {4},
location = {Szeged, Hungary},
series = {ESEC/FSE '11}
}

@inproceedings{wei2022chain,
 author = {Wei, Jason and Wang, Xuezhi and Schuurmans, Dale and Bosma, Maarten and ichter, brian and Xia, Fei and Chi, Ed and Le, Quoc V and Zhou, Denny},
 booktitle = {Advances in Neural Information Processing Systems},
 editor = {S. Koyejo and S. Mohamed and A. Agarwal and D. Belgrave and K. Cho and A. Oh},
 pages = {24824--24837},
 publisher = {Curran Associates, Inc.},
 title = {Chain-of-Thought Prompting Elicits Reasoning in Large Language Models},
 url = {https://proceedings.neurips.cc/paper_files/paper/2022/file/9d5609613524ecf4f15af0f7b31abca4-Paper-Conference.pdf},
 volume = {35},
 year = {2022},
 doi = {10.52202/068431-1800}
}

@misc{deepseekai2025deepseekv32pushingfrontieropen,
      title={DeepSeek-V3.2: Pushing the Frontier of Open Large Language Models}, 
      author={DeepSeek-AI and Aixin Liu and Aoxue Mei and Bangcai Lin and others},
      year={2025},
      eprint={2512.02556},
      archivePrefix={arXiv},
      primaryClass={cs.CL},
      url={https://arxiv.org/abs/2512.02556},
      doi={10.48550/arXiv.2512.02556},
}

@misc{openai2025gptoss120bgptoss20bmodel,
      title={gpt-oss-120b and gpt-oss-20b Model Card}, 
      author={OpenAI and Sandhini Agarwal and Lama Ahmad and others},
      year={2025},
      eprint={2508.10925},
      archivePrefix={arXiv},
      primaryClass={cs.CL},
      url={https://arxiv.org/abs/2508.10925},
      doi={10.48550/arXiv.2508.10925},
}

@article{shi2024code,
  title={From code to correctness: Closing the last mile of code generation with hierarchical debugging},
  author={Shi, Yuling and Wang, Songsong and Wan, Chengcheng and Wang, Min and Gu, Xiaodong},
  journal={arXiv preprint arXiv:2410.01215},
  year={2024},
  doi={10.48550/arXiv.2410.01215}
}

@article{jia2009higher,
title = {Higher Order Mutation Testing},
journal = {Information and Software Technology},
volume = {51},
number = {10},
pages = {1379-1393},
year = {2009},
note = {Source Code Analysis and Manipulation, SCAM 2008},
issn = {0950-5849},
doi = {https://doi.org/10.1016/j.infsof.2009.04.016},
url = {https://www.sciencedirect.com/science/article/pii/S0950584909000688},
author = {Yue Jia and Mark Harman}
}

@article{OFFUT92COUPLING,
author = {Offutt, A. Jefferson},
title = {Investigations of the software testing coupling effect},
year = {1992},
issue_date = {Jan. 1992},
publisher = {Association for Computing Machinery},
address = {New York, NY, USA},
volume = {1},
number = {1},
issn = {1049-331X},
url = {https://doi.org/10.1145/125489.125473},
doi = {10.1145/125489.125473},
journal = {ACM Trans. Softw. Eng. Methodol.},
month = jan,
pages = {5–20},
numpages = {16}
}

@inproceedings{beller2015much,
  author={Beller, Moritz and Gousios, Georgios and Zaidman, Andy},
  booktitle={2015 IEEE/ACM 37th IEEE International Conference on Software Engineering}, 
  title={How (Much) Do Developers Test?}, 
  year={2015},
  volume={2},
  number={},
  pages={559-562},
  doi={10.1109/ICSE.2015.193}}

@inproceedings{QUICKCHECK,
author = {Claessen, Koen and Hughes, John},
title = {QuickCheck: a lightweight tool for random testing of Haskell programs},
year = {2000},
isbn = {1581132026},
publisher = {Association for Computing Machinery},
address = {New York, NY, USA},
url = {https://doi.org/10.1145/351240.351266},
doi = {10.1145/351240.351266},
booktitle = {Proceedings of the Fifth ACM SIGPLAN International Conference on Functional Programming},
pages = {268–279},
numpages = {12},
series = {ICFP '00}
}

@inproceedings{KLEE,
author = {Cadar, Cristian and Dunbar, Daniel and Engler, Dawson},
title = {KLEE: unassisted and automatic generation of high-coverage tests for complex systems programs},
year = {2008},
publisher = {USENIX Association},
address = {USA},
booktitle = {Proceedings of the 8th USENIX Conference on Operating Systems Design and Implementation},
pages = {209–224},
numpages = {16},
location = {San Diego, California},
series = {OSDI'08},
}

@inproceedings{deljouyi2024leveraging,
  author={Deljouyi, Amirhossein and Koohestani, Roham and Izadi, Maliheh and Zaidman, Andy},
  booktitle={2025 IEEE/ACM 47th International Conference on Software Engineering (ICSE)}, 
  title={Leveraging Large Language Models for Enhancing the Understandability of Generated Unit Tests}, 
  year={2025},
  volume={},
  number={},
  pages={1449-1461},
  doi={10.1109/ICSE55347.2025.00032}}

@article{maciver2019hypothesis, doi = {10.21105/joss.01891}, url = {https://doi.org/10.21105/joss.01891}, year = {2019}, publisher = {The Open Journal}, volume = {4}, number = {43}, pages = {1891}, author = {MacIver, David R. and Hatfield-Dodds, Zac and Contributors, Many Other}, title = {Hypothesis: A new approach to property-based testing}, journal = {Journal of Open Source Software} }

@inproceedings{yang2019refactory,
  author={Hu, Yang and Ahmed, Umair Z. and Mechtaev, Sergey and Leong, Ben and Roychoudhury, Abhik},
  booktitle={2019 34th IEEE/ACM International Conference on Automated Software Engineering (ASE)}, 
  title={Re-Factoring Based Program Repair Applied to Programming Assignments}, 
  year={2019},
  volume={},
  number={},
  pages={388-398},
  doi={10.1109/ASE.2019.00044}}

@article{yang2024whitefox,
author = {Yang, Chenyuan and Deng, Yinlin and Lu, Runyu and Yao, Jiayi and Liu, Jiawei and Jabbarvand, Reyhaneh and Zhang, Lingming},
title = {WhiteFox: White-Box Compiler Fuzzing Empowered by Large Language Models},
year = {2024},
issue_date = {October 2024},
publisher = {Association for Computing Machinery},
address = {New York, NY, USA},
volume = {8},
number = {OOPSLA2},
url = {https://doi.org/10.1145/3689736},
doi = {10.1145/3689736},
journal = {Proc. ACM Program. Lang.},
month = oct,
articleno = {296},
numpages = {27}
}

@article{ArcuriYao2014,
title = {Co-evolutionary automatic programming for software development},
journal = {Information Sciences},
volume = {259},
pages = {412-432},
year = {2014},
issn = {0020-0255},
doi = {https://doi.org/10.1016/j.ins.2009.12.019},
url = {https://www.sciencedirect.com/science/article/pii/S0020025509005490},
author = {Andrea Arcuri and Xin Yao}
}

@inproceedings{CodeDefenders2019,
author = {Fraser, Gordon and Gambi, Alessio and Kreis, Marvin and Rojas, Jos\'{e} Miguel},
title = {Gamifying a Software Testing Course with Code Defenders},
year = {2019},
isbn = {9781450358903},
publisher = {Association for Computing Machinery},
address = {New York, NY, USA},
url = {https://doi.org/10.1145/3287324.3287471},
doi = {10.1145/3287324.3287471},
booktitle = {Proceedings of the 50th ACM Technical Symposium on Computer Science Education},
pages = {571–577},
numpages = {7},
location = {Minneapolis, MN, USA},
series = {SIGCSE '19}
}

@inproceedings{PeX,
author="Tillmann, Nikolai
and de Halleux, Jonathan",
editor="Beckert, Bernhard
and H{\"a}hnle, Reiner",
title="Pex--White Box Test Generation for .NET",
booktitle="Tests and Proofs",
year="2008",
publisher="Springer Berlin Heidelberg",
address="Berlin, Heidelberg",
pages="134--153",
isbn="978-3-540-79124-9",
doi="https://doi.org/10.1007/978-3-540-79124-9_10"
}

@article{KORAT,
author = {Boyapati, Chandrasekhar and Khurshid, Sarfraz and Marinov, Darko},
title = {Korat: automated testing based on Java predicates},
year = {2002},
issue_date = {July 2002},
publisher = {Association for Computing Machinery},
address = {New York, NY, USA},
volume = {27},
number = {4},
issn = {0163-5948},
url = {https://doi.org/10.1145/566171.566191},
doi = {10.1145/566171.566191},
journal = {SIGSOFT Softw. Eng. Notes},
month = jul,
pages = {123–133},
numpages = {11}
}

@inproceedings{beller2015and,
author = {Beller, Moritz and Gousios, Georgios and Panichella, Annibale and Zaidman, Andy},
title = {When, how, and why developers (do not) test in their IDEs},
year = {2015},
isbn = {9781450336758},
publisher = {Association for Computing Machinery},
address = {New York, NY, USA},
url = {https://doi.org/10.1145/2786805.2786843},
doi = {10.1145/2786805.2786843},
booktitle = {Proceedings of the 2015 10th Joint Meeting on Foundations of Software Engineering},
pages = {179–190},
numpages = {12},
location = {Bergamo, Italy},
series = {ESEC/FSE 2015}
}

@inproceedings{fdr,
  author={Rothermel, G. and Untch, R.H. and Chengyun Chu and Harrold, M.J.},
  booktitle={Proceedings IEEE International Conference on Software Maintenance - 1999 (ICSM'99). 'Software Maintenance for Business Change' (Cat. No.99CB36360)}, 
  title={Test case prioritization: an empirical study}, 
  year={1999},
  volume={},
  number={},
  pages={179-188},
  doi={10.1109/ICSM.1999.792604}}

@article{chen2021codex,
  title={Evaluating Large Language Models Trained on Code},
  author={Mark Chen and Jerry Tworek and Heewoo Jun and others},
  year={2021},
  eprint={2107.03374},
  archivePrefix={arXiv},
  primaryClass={cs.LG},
  doi={10.48550/arXiv.2107.03374}
}

@article{liu2024deepseek,
  title={Deepseek-v3 technical report},
  author={Liu, Aixin and Feng, Bei and Xue, Bing and Wang, Bingxuan and Wu, Bochao and Lu, Chengda and Zhao, Chenggang and Deng, Chengqi and Zhang, Chenyu and Ruan, Chong and others},
  journal={arXiv preprint arXiv:2412.19437},
  year={2024},
  doi={10.48550/arXiv.2412.19437}
}

@article{testforge,
  title={TestForge: Feedback-Driven, Agentic Test Suite Generation},
  author={Jain, Kush and Goues, Claire Le},
  journal={arXiv preprint arXiv:2503.14713},
  year={2025},
  doi={10.48550/arXiv.2503.14713}
}

@inproceedings{hemmati2015effective,
  author={Hemmati, Hadi},
  booktitle={2015 IEEE International Conference on Software Quality, Reliability and Security}, 
  title={How Effective Are Code Coverage Criteria?}, 
  year={2015},
  volume={},
  number={},
  pages={151-156},
  doi={10.1109/QRS.2015.30}}

@inproceedings{gopinath2014code,
author = {Gopinath, Rahul and Jensen, Carlos and Groce, Alex},
title = {Code coverage for suite evaluation by developers},
year = {2014},
isbn = {9781450327565},
publisher = {Association for Computing Machinery},
address = {New York, NY, USA},
url = {https://doi.org/10.1145/2568225.2568278},
doi = {10.1145/2568225.2568278},
booktitle = {Proceedings of the 36th International Conference on Software Engineering},
pages = {72–82},
numpages = {11},
location = {Hyderabad, India},
series = {ICSE 2014}
}

@inproceedings{cai2005effect,
author = {Cai, Xia and Lyu, Michael R.},
title = {The effect of code coverage on fault detection under different testing profiles},
year = {2005},
isbn = {1595931155},
publisher = {Association for Computing Machinery},
address = {New York, NY, USA},
url = {https://doi.org/10.1145/1083274.1083288},
doi = {10.1145/1083274.1083288},
booktitle = {Proceedings of the 1st International Workshop on Advances in Model-Based Testing},
pages = {1–7},
numpages = {7},
location = {St. Louis, Missouri},
series = {A-MOST '05}
}

@misc{cobertura-tool,
  author       = {{The Cobertura Team}},
  title        = {Cobertura},
    year       = {2015},
  howpublished = {\url{https://github.com/cobertura/cobertura}}
 
}

@misc{replication,
  author       = {Pengyu Chang and Yixiong Fang and Silin Chen and Yuling Shi and Beijun Shen and Xiaodong Gu},
  title        = {The replication package},
  year         = {2025},
  howpublished = {\url{https://github.com/jmueducn/AdverTest}}
}

@book{jorgensen2013software,
  title={Software testing: a craftsman's approach},
  author={Jorgensen, Paul C},
  year={2013},
  publisher={Auerbach Publications},
  doi = {https://doi.org/10.1201/9781439889503}
}

@inproceedings{pynguin,
author = {Lukasczyk, Stephan and Fraser, Gordon},
title = {Pynguin: automated unit test generation for Python},
year = {2022},
isbn = {9781450392235},
publisher = {Association for Computing Machinery},
address = {New York, NY, USA},
url = {https://doi.org/10.1145/3510454.3516829},
doi = {10.1145/3510454.3516829},
booktitle = {Proceedings of the ACM/IEEE 44th International Conference on Software Engineering: Companion Proceedings},
pages = {168–172},
numpages = {5},
location = {Pittsburgh, Pennsylvania},
series = {ICSE '22}
}

@inproceedings{defects4j,
author = {Just, Ren\'{e} and Jalali, Darioush and Ernst, Michael D.},
title = {Defects4J: a database of existing faults to enable controlled testing studies for Java programs},
year = {2014},
isbn = {9781450326452},
publisher = {Association for Computing Machinery},
address = {New York, NY, USA},
url = {https://doi.org/10.1145/2610384.2628055},
doi = {10.1145/2610384.2628055},
booktitle = {Proceedings of the 2014 International Symposium on Software Testing and Analysis},
pages = {437–440},
numpages = {4},
location = {San Jose, CA, USA},
series = {ISSTA 2014}
}

@INPROCEEDINGS {GrowingBugsICSE21, author = {Yanjie Jiang and Hui Liu and Nan Niu and Lu Zhang and Yamin Hu}, booktitle = {IEEE/ACM 43rd International Conference on Software Engineering (ICSE 2021)}, title = {Extracting Concise Bug-Fixing Patches from Human-Written Patches in Version Control Systems}, year = {2021}, pages = {686-698}, doi = {10.1109/ICSE43902.2021.00069}, url = {https://doi.ieeecomputersociety.org/10.1109/ICSE43902.2021.00069}, publisher = {IEEE Computer Society}, address = {Los Alamitos, CA, USA}, month = {may} }

@ARTICLE{GrowingBugsTSE2022,   author={Jiang, Yanjie and Liu, Hui and Luo, Xiaoqing and Zhu, Zhihao and Chi, Xiaye and Niu, Nan and Zhang, Yuxia and Hu, Yamin and Bian, Pan and Zhang, Lu},   journal={IEEE Transactions on Software Engineering},    title={BugBuilder: An Automated Approach to Building Bug Repository},    year={2022},   volume={},   number={},   pages={1-22},   doi={10.1109/TSE.2022.3177713}}

@inproceedings{NaturalnessOfBugsFSE2022, author = {Jiang, Yanjie and Liu, Hui and Zhang, Yuxia and Ji, Weixing and Zhong, Hao and Zhang, Lu}, title = {Do Bugs Lead to Unnaturalness of Source Code?}, year = {2022}, isbn = {9781450394130}, publisher = {Association for Computing Machinery}, address = {New York, NY, USA}, url = {https://doi.org/10.1145/3540250.3549149}, doi = {10.1145/3540250.3549149}, booktitle = {Proceedings of the 30th ACM Joint European Software Engineering Conference and Symposium on the Foundations of Software Engineering}, pages = {1085–1096}, numpages = {12}, location = {Singapore, Singapore}, series = {ESEC/FSE 2022} }

@article{mutap,
title = {Effective test generation using pre-trained Large Language Models and mutation testing},
journal = {Information and Software Technology},
volume = {171},
pages = {107468},
year = {2024},
issn = {0950-5849},
doi = {https://doi.org/10.1016/j.infsof.2024.107468},
url = {https://www.sciencedirect.com/science/article/pii/S0950584924000739},
author = {Arghavan Moradi Dakhel and Amin Nikanjam and Vahid Majdinasab and Foutse Khomh and Michel C. Desmarais}
}

@inproceedings{hits,
author = {Wang, Zejun and Liu, Kaibo and Li, Ge and Jin, Zhi},
title = {HITS: High-coverage LLM-based Unit Test Generation via Method Slicing},
year = {2024},
isbn = {9798400712487},
publisher = {Association for Computing Machinery},
address = {New York, NY, USA},
url = {https://doi.org/10.1145/3691620.3695501},
doi = {10.1145/3691620.3695501},
booktitle = {Proceedings of the 39th IEEE/ACM International Conference on Automated Software Engineering},
pages = {1258–1268},
numpages = {11},
location = {Sacramento, CA, USA},
series = {ASE '24}
}

@inproceedings{chatunitest,
author = {Chen, Yinghao and Hu, Zehao and Zhi, Chen and Han, Junxiao and Deng, Shuiguang and Yin, Jianwei},
title = {ChatUniTest: A Framework for LLM-Based Test Generation},
year = {2024},
isbn = {9798400706585},
publisher = {Association for Computing Machinery},
address = {New York, NY, USA},
url = {https://doi.org/10.1145/3663529.3663801},
doi = {10.1145/3663529.3663801},
booktitle = {Companion Proceedings of the 32nd ACM International Conference on the Foundations of Software Engineering},
pages = {572–576},
numpages = {5},
location = {Porto de Galinhas, Brazil},
series = {FSE 2024}
}

@inproceedings{Pacheco2007,
author = {Pacheco, Carlos and Ernst, Michael D.},
title = {Randoop: feedback-directed random testing for Java},
year = {2007},
isbn = {9781595938657},
publisher = {Association for Computing Machinery},
address = {New York, NY, USA},
url = {https://doi.org/10.1145/1297846.1297902},
doi = {10.1145/1297846.1297902},
booktitle = {Companion to the 22nd ACM SIGPLAN Conference on Object-Oriented Programming Systems and Applications Companion},
pages = {815–816},
numpages = {2},
location = {Montreal, Quebec, Canada},
series = {OOPSLA '07}
}

@article{Tufano2020,
  title={Unit test case generation with transformers and focal context},
  author={Tufano, Michele and Drain, Dawn and Svyatkovskiy, Alexey and Deng, Shao Kun and Sundaresan, Neel},
  journal={arXiv preprint arXiv:2009.05617},
  year={2020},
  doi={10.48550/arXiv.2009.05617}
}

@inproceedings{Quixbugs,
author = {Lin, Derrick and Koppel, James and Chen, Angela and Solar-Lezama, Armando},
title = {QuixBugs: a multi-lingual program repair benchmark set based on the quixey challenge},
year = {2017},
isbn = {9781450355148},
publisher = {Association for Computing Machinery},
address = {New York, NY, USA},
url = {https://doi.org/10.1145/3135932.3135941},
doi = {10.1145/3135932.3135941},
booktitle = {Proceedings Companion of the 2017 ACM SIGPLAN International Conference on Systems, Programming, Languages, and Applications: Software for Humanity},
pages = {55–56},
numpages = {2},
location = {Vancouver, BC, Canada},
series = {SPLASH Companion 2017}
}

@article{Alagarsamy2024,
title = {A3Test: Assertion-Augmented Automated Test case generation},
journal = {Information and Software Technology},
volume = {176},
pages = {107565},
year = {2024},
issn = {0950-5849},
doi = {https://doi.org/10.1016/j.infsof.2024.107565},
url = {https://www.sciencedirect.com/science/article/pii/S0950584924001708},
author = {Saranya Alagarsamy and Chakkrit Tantithamthavorn and Aldeida Aleti}
}

@inproceedings{Villmow2021,
    title = "{C}on{T}est: A Unit Test Completion Benchmark featuring Context",
    author = "Villmow, Johannes  and
      Depoix, Jonas  and
      Ulges, Adrian",
    editor = "Lachmy, Royi  and
      Yao, Ziyu  and
      Durrett, Greg  and
      Gligoric, Milos  and
      Li, Junyi Jessy  and
      Mooney, Ray  and
      Neubig, Graham  and
      Su, Yu  and
      Sun, Huan  and
      Tsarfaty, Reut",
    booktitle = "Proceedings of the 1st Workshop on Natural Language Processing for Programming (NLP4Prog 2021)",
    month = aug,
    year = "2021",
    address = "Online",
    publisher = "Association for Computational Linguistics",
    url = "https://aclanthology.org/2021.nlp4prog-1.2/",
    doi = "10.18653/v1/2021.nlp4prog-1.2",
    pages = "17--25"
}

@inproceedings{Nie2023,
  author={Nie, Pengyu and Banerjee, Rahul and Li, Junyi Jessy and Mooney, Raymond J. and Gligoric, Milos},
  booktitle={2023 IEEE/ACM 45th International Conference on Software Engineering (ICSE)}, 
  title={Learning Deep Semantics for Test Completion}, 
  year={2023},
  volume={},
  number={},
  pages={2111-2123},
  doi={10.1109/ICSE48619.2023.00178}}

@inproceedings{barboni2025smartmutant,
  author={Barboni, Morena and Lampa, Filippo and Morichetta, Andrea and Polini, Andrea and Zulkoski, Edward},
  booktitle={2025 IEEE International Conference on Artificial Intelligence Testing (AITest)}, 
  title={Mutant-Driven Test Generation for Ethereum Smart Contracts via LLMs}, 
  year={2025},
  volume={},
  number={},
  pages={209-216},
  doi={10.1109/AITest66680.2025.00033}}

@inproceedings{mutationatmeta,
author = {Harman, Mark and Ritchey, Jillian and Harper, Inna and Sengupta, Shubho and Mao, Ke and Gulati, Abhishek and Foster, Christopher and Robert, Herv\'{e}},
title = {Mutation-Guided LLM-based Test Generation at Meta},
year = {2025},
isbn = {9798400712760},
publisher = {Association for Computing Machinery},
address = {New York, NY, USA},
url = {https://doi.org/10.1145/3696630.3728544},
doi = {10.1145/3696630.3728544},
booktitle = {Proceedings of the 33rd ACM International Conference on the Foundations of Software Engineering},
pages = {180–191},
numpages = {12},
location = {Clarion Hotel Trondheim, Trondheim, Norway},
series = {FSE Companion '25}
}

@article{coverup,
author = {Altmayer Pizzorno, Juan and Berger, Emery D.},
title = {CoverUp: Effective High Coverage Test Generation for Python},
year = {2025},
issue_date = {July 2025},
publisher = {Association for Computing Machinery},
address = {New York, NY, USA},
volume = {2},
number = {FSE},
url = {https://doi.org/10.1145/3729398},
doi = {10.1145/3729398},
journal = {Proc. ACM Softw. Eng.},
month = jun,
articleno = {FSE128},
numpages = {23}
}

@inproceedings{Rao2023,
  author={Rao, Nikitha and Jain, Kush and Alon, Uri and Goues, Claire Le and Hellendoorn, Vincent J.},
  booktitle={2023 38th IEEE/ACM International Conference on Automated Software Engineering (ASE)}, 
  title={CAT-LM Training Language Models on Aligned Code And Tests}, 
  year={2023},
  volume={},
  number={},
  pages={409-420},
  doi={10.1109/ASE56229.2023.00193}}

@inproceedings{Siddiqa2023,
author = {Siddiq, Mohammed Latif and Da Silva Santos, Joanna Cecilia and Tanvir, Ridwanul Hasan and Ulfat, Noshin and Al Rifat, Fahmid and Carvalho Lopes, Vin{\'i}cius},
title = {Using Large Language Models to Generate JUnit Tests: An Empirical Study},
year = {2024},
isbn = {9798400717017},
publisher = {Association for Computing Machinery},
address = {New York, NY, USA},
url = {https://doi.org/10.1145/3661167.3661216},
doi = {10.1145/3661167.3661216},
booktitle = {Proceedings of the 28th International Conference on Evaluation and Assessment in Software Engineering},
pages = {313–322},
numpages = {10},
location = {Salerno, Italy},
series = {EASE '24}
}

@article{Schafer2023,
  author={Schäfer, Max and Nadi, Sarah and Eghbali, Aryaz and Tip, Frank},
  journal={IEEE Transactions on Software Engineering}, 
  title={An Empirical Evaluation of Using Large Language Models for Automated Unit Test Generation}, 
  year={2024},
  volume={50},
  number={1},
  pages={85-105},
  doi={10.1109/TSE.2023.3334955}}

@inproceedings{Lemieux2023,
  author={Lemieux, Caroline and Inala, Jeevana Priya and Lahiri, Shuvendu K. and Sen, Siddhartha},
  booktitle={2023 IEEE/ACM 45th International Conference on Software Engineering (ICSE)}, 
  title={CodaMosa: Escaping Coverage Plateaus in Test Generation with Pre-trained Large Language Models}, 
  year={2023},
  volume={},
  number={},
  pages={919-931},
  doi={10.1109/ICSE48619.2023.00085}}

@inproceedings{Alshahwan2024,
author = {Alshahwan, Nadia and Chheda, Jubin and Finogenova, Anastasia and Gokkaya, Beliz and Harman, Mark and Harper, Inna and Marginean, Alexandru and Sengupta, Shubho and Wang, Eddy},
title = {Automated Unit Test Improvement using Large Language Models at Meta},
year = {2024},
isbn = {9798400706585},
publisher = {Association for Computing Machinery},
address = {New York, NY, USA},
url = {https://doi.org/10.1145/3663529.3663839},
doi = {10.1145/3663529.3663839},
booktitle = {Companion Proceedings of the 32nd ACM International Conference on the Foundations of Software Engineering},
pages = {185–196},
numpages = {12},
location = {Porto de Galinhas, Brazil},
series = {FSE 2024}
}

@article{DeMillo1978,
  author={DeMillo, R.A. and Lipton, R.J. and Sayward, F.G.},
  journal={Computer}, 
  title={Hints on Test Data Selection: Help for the Practicing Programmer}, 
  year={1978},
  volume={11},
  number={4},
  pages={34-41},
  doi={10.1109/C-M.1978.218136}}

@inproceedings{straubinger2025mutation,
  author={Straubinger, Philipp and Kreis, Marvin and Lukasczyk, Stephan and Fraser, Gordon},
  booktitle={2025 IEEE International Conference on Software Testing, Verification and Validation Workshops (ICSTW)}, 
  title={Mutation Testing via Iterative Large Language Model-Driven Scientific Debugging}, 
  year={2025},
  volume={},
  number={},
  pages={358-367},
  doi={10.1109/ICSTW64639.2025.10962485}}

@inproceedings{Coles2016,
author = {Coles, Henry and Laurent, Thomas and Henard, Christopher and Papadakis, Mike and Ventresque, Anthony},
title = {PIT: a practical mutation testing tool for Java (demo)},
year = {2016},
isbn = {9781450343909},
publisher = {Association for Computing Machinery},
address = {New York, NY, USA},
url = {https://doi.org/10.1145/2931037.2948707},
doi = {10.1145/2931037.2948707},
booktitle = {Proceedings of the 25th International Symposium on Software Testing and Analysis},
pages = {449–452},
numpages = {4},
location = {Saarbr\"{u}cken, Germany},
series = {ISSTA 2016}
}

@inproceedings{Just2011,
  author={Just, René and Schweiggert, Franz and Kapfhammer, Gregory M.},
  booktitle={2011 26th IEEE/ACM International Conference on Automated Software Engineering (ASE 2011)}, 
  title={MAJOR: An efficient and extensible tool for mutation analysis in a Java compiler}, 
  year={2011},
  volume={},
  number={},
  pages={612-615},
  doi={10.1109/ASE.2011.6100138}}

@inproceedings{Degiovanni2022,
  author={Degiovanni, Renzo and Papadakis, Mike},
  booktitle={2022 IEEE International Conference on Software Testing, Verification and Validation Workshops (ICSTW)}, 
  title={µBert: Mutation Testing using Pre-Trained Language Models}, 
  year={2022},
  volume={},
  number={},
  pages={160-169},
  doi={10.1109/ICSTW55395.2022.00039}}

@article{Tip2025,
  author={Tip, Frank and Bell, Jonathan and Schäfer, Max},
  journal={IEEE Transactions on Software Engineering}, 
  title={LLMorpheus: Mutation Testing Using Large Language Models}, 
  year={2025},
  volume={51},
  number={6},
  pages={1645-1665},
  doi={10.1109/TSE.2025.3562025}}

@article{WangMutation2024,
author = {Wang, Bo and Chen, Mingda and Deng, Ming and Lin, Youfang and Harman, Mark and Papadakis, Mike and Zhang, Jie M.},
title = {A Comprehensive Study on Large Language Models for Mutation Testing},
year = {2026},
publisher = {Association for Computing Machinery},
address = {New York, NY, USA},
issn = {1049-331X},
url = {https://doi.org/10.1145/3805038},
doi = {10.1145/3805038},
note = {Just Accepted},
journal = {ACM Trans. Softw. Eng. Methodol.},
month = mar
}

@inproceedings{Garg2024,
  author={Garg, Aayush and Degiovanni, Renzo and Papadakis, Mike and Traon, Yves Le},
  booktitle={2024 IEEE Conference on Software Testing, Verification and Validation (ICST)}, 
  title={On the Coupling between Vulnerabilities and LLM-Generated Mutants: A Study on Vul4J Dataset}, 
  year={2024},
  volume={},
  number={},
  pages={305-316},
  doi={10.1109/ICST60714.2024.00035}}

@inproceedings{badertdinov2026swe,
 author = {Badertdinov, Ibragim and Golubev, Alexander and Nekrashevich, Maksim and Shevtsov, Anton and Karasik, Simon and Andriushchenko, Andrei and Trofimova, Maria and Litvintseva, Daria and Yangel, Boris},
 booktitle = {Advances in Neural Information Processing Systems},
 editor = {D. Belgrave and C. Zhang and H. Lin and R. Pascanu and P. Koniusz and M. Ghassemi and N. Chen},
 pages = {},
 publisher = {Curran Associates, Inc.},
 title = {{SWE-Rebench}: An Automated Pipeline for Task Collection and Decontaminated Evaluation of Software Engineering Agents},
 url = {https://proceedings.neurips.cc/paper_files/paper/2025/file/21bec6ace947b1b58967b945c8ac0f10-Paper-Datasets_and_Benchmarks_Track.pdf},
 volume = {38},
 year = {2025},
 doi = {10.52202/085713-0788}
}

@article{chen2025swe,
  title={Swe-exp: Experience-driven software issue resolution},
  author={Chen, Silin and Lin, Shaoxin and Shi, Yuling and Lian, Heng and Gu, Xiaodong and Yun, Longfei and Chen, Dong and Cao, Lin and Liu, Jiyang and Xia, Nu and others},
  journal={arXiv preprint arXiv:2507.23361},
  year={2025},
  doi={10.48550/arXiv.2507.23361}
}

@article{lin2024llms,
  title={Llms as continuous learners: Improving the reproduction of defective code in software issues},
  author={Lin, Yalan and Ma, Yingwei and Cao, Rongyu and Li, Binhua and Huang, Fei and Gu, Xiaodong and Li, Yongbin},
  journal={arXiv preprint arXiv:2411.13941},
  year={2024},
  doi={10.48550/arXiv.2411.13941}
}

@article{202605.2065,
	doi = {10.20944/preprints202605.2065.v1},
	url = {https://doi.org/10.20944/preprints202605.2065.v1},
	year = 2026,
	month = {May},
	publisher = {Preprints},
	author = {Yifei Wang and Ziteng Wang and Yuling Shi and Silin Chen and Xinrui Wang and Yueqi Wang and Beijun Shen and Linjing Li and Xiaodong Gu and Julian McAuley and Daniel Dajun Zeng},
	title = {Context Compression for LLM Agents: A Survey of Methods, Failure Modes, and Evaluation},
	journal = {Preprints}
}

@INPROCEEDINGS{shi2025between,
  author={Shi, Yuling and Zhang, Hongyu and Wan, Chengcheng and Gu, Xiaodong},
  booktitle={2025 IEEE/ACM 47th International Conference on Software Engineering (ICSE)}, 
  title={Between Lines of Code: Unraveling the Distinct Patterns of Machine and Human Programmers}, 
  year={2025},
  volume={},
  number={},
  pages={1628-1639},
  doi={10.1109/ICSE55347.2025.00005}}

@article{li2025swe,
  title={Swe-debate: Competitive multi-agent debate for software issue resolution},
  author={Li, Han and Shi, Yuling and Lin, Shaoxin and Gu, Xiaodong and Lian, Heng and Wang, Xin and Jia, Yantao and Huang, Tao and Wang, Qianxiang},
  journal={arXiv preprint arXiv:2507.23348},
  year={2025},
  doi={10.48550/arXiv.2507.23348}
}

@inproceedings{shi2025longcodezip,
  author={Shi, Yuling and Qian, Yichun and Zhang, Hongyu and Shen, Beijun and Gu, Xiaodong},
  booktitle={2025 40th IEEE/ACM International Conference on Automated Software Engineering (ASE)}, 
  title={LongCodeZip: Compress Long Context for Code Language Models}, 
  year={2025},
  volume={},
  number={},
  pages={141-153},
  doi={10.1109/ASE63991.2025.00020}}

@article{wang2026swe,
  title={SWE-Pruner: Self-Adaptive Context Pruning for Coding Agents},
  author={Wang, Yuhang and Shi, Yuling and Yang, Mo and Zhang, Rongrui and He, Shilin and Lian, Heng and Chen, Yuting and Ye, Siyu and Cai, Kai and Gu, Xiaodong},
  journal={arXiv preprint arXiv:2601.16746},
  year={2026},
  doi={10.48550/arXiv.2601.16746}
}

@article{hu2026line,
author = {Hu, Chao and Zeng, Wenhao and Shi, Yuling and Shen, Beijun and Gu, Xiaodong},
title = {In Line with Context: Repository-Level Code Generation via Context Inlining},
year = {2026},
issue_date = {July 2026},
publisher = {Association for Computing Machinery},
address = {New York, NY, USA},
volume = {3},
number = {FSE},
url = {https://doi.org/10.1145/3797094},
doi = {10.1145/3797094},
journal = {Proc. ACM Softw. Eng.},
month = jun,
articleno = {FSE066},
numpages = {23}
}

@inproceedings{zeng2026glimprouter,
    title = "{G}limp{R}outer: Efficient Collaborative Inference by Glimpsing One Token of Thoughts",
    author = "Zeng, Wenhao  and
      Zhang, Xuteng  and
      Shi, Yuling  and
      Hu, Chao  and
      Chen, Yuting  and
      Shen, Beijun  and
      Gu, Xiaodong",
    editor = "Liakata, Maria  and
      Moreira, Viviane P.  and
      Zhang, Jiajun  and
      Jurgens, David",
    booktitle = "Findings of the {A}ssociation for {C}omputational {L}inguistics: {ACL} 2026",
    month = jul,
    year = "2026",
    address = "San Diego, California, United States",
    publisher = "Association for Computational Linguistics",
    url = "https://aclanthology.org/2026.findings-acl.885/",
    doi = "10.18653/v1/2026.findings-acl.885",
    pages = "17850--17864",
    ISBN = "979-8-89176-395-1"
}

\end{document}